\documentclass[11pt]{article}
\usepackage{amsfonts}
\usepackage{amsmath,amssymb}
\usepackage{bbm}
\usepackage{bm}
\usepackage{color}
\usepackage{slashed}
\usepackage{cite}
\usepackage[utf8]{inputenc}
\usepackage{graphicx}
\usepackage{bbold}
\DeclareGraphicsExtensions{.pdf,.png,.jpg}
\newcommand{\rmi}[1]{{\mbox{\scriptsize #1}}}

\newcommand{\lnfplus}{\ln_\rmi{f}^+}
\newcommand{\lnfminus}{\ln_\rmi{f}^-}
\newcommand{\lnbplus}{\ln_\rmi{b}^+}
\newcommand{\lnbminus}{\ln_\rmi{b}^-}
\newcommand{\lifplus}[1]{{\mbox{li}}_\rmi{${#1}$f}^+}
\newcommand{\lifminus}[1]{{\mbox{li}}_\rmi{${#1}$f}^-}
\newcommand{\libplus}[1]{{\mbox{li}}_\rmi{${#1}$b}^+}
\newcommand{\libminus}[1]{{\mbox{li}}_\rmi{${#1}$b}^-}

\usepackage[colorlinks=false,citecolor=cyan,urlcolor=blue,bookmarks=true,bookmarks=true,bookmarksopen=true,bookmarksnumbered=true,bookmarksopenlevel=3]{hyperref}

\definecolor{airforceblue}{rgb}{0.36, 0.54, 0.66}
\definecolor{steelblue}{rgb}{0.27, 0.51, 0.71}
\definecolor{amber}{rgb}{1.0, 0.49, 0.0}

\def\simg{{\ \lower-1.2pt\vbox{\hbox{\rlap{$>$}\lower6pt\vbox{\hbox{$\sim$}}}}\ }}
\def\siml{{\ \lower-1.2pt\vbox{\hbox{\rlap{$<$}\lower6pt\vbox{\hbox{$\sim$}}}}\ }}

\makeatletter \@addtoreset{equation}{section} \makeatother

\begin{document}

\flushbottom

\begin{titlepage}

\begin{centering}

\vfill

{\Large{\bf
 Freeze-in produced dark matter  in the 
 \\
 ultra-relativistic regime
}} 

\vspace{0.8cm}

S.~Biondini$^a$ and J.~Ghiglieri$^b$

\vspace{0.8cm}

 $^\rmi{a}$%
{\em Department of Physics, University of Basel,
\\
 Klingelbergstr. 82, CH-4056 Basel, Switzerland} 
\\
\vspace{0.25 cm}
 $^\rmi{b}$%
{\em SUBATECH, Universit\'e de Nantes, IMT Atlantique, IN2P3/CNRS,
\\
4 rue Alfred Kastler, La Chantrerie BP 20722, 44307 Nantes, France}
\vspace*{0.8cm}

\end{centering}

\vspace*{0.5cm}
 
\noindent
\textbf{Abstract}: When dark matter particles only feebly interact with plasma constituents in the early universe, they never reach thermal equilibrium. As opposed to the freeze-out mechanism, where the dark matter abundance is determined at $T \ll M$,
the energy density of a feebly interacting state builds up and increases over $T \simg M$. In this work, we address the impact of the high-temperature regime on the dark matter production rate, where the dark and Standard Model particles are ultra-relativistic and nearly light-like. In this setting, multiple soft scatterings, as well as $2 \to 2$ processes, are found to give a large contribution to the production rate. Within the model we consider in this work, namely a Majorana fermion dark matter of mass $M$ accompanied  by a heavier scalar --- with mass splitting $\Delta M$ ---  which shares interactions with the visible sector, the energy density can be dramatically underestimated when neglecting the high-temperature dynamics.  We find that the overall effective $1 \leftrightarrow 2$ and $2 \to2$ high-temperature contributions to dark-matter production give $\mathcal{O}(10)$ (20\%) corrections  for $\Delta M /M =0.1$ ($\Delta M /M =10$) to the  Born production rate with in-vacuum masses and matrix elements.
We also assess the impact of bound-state effects on the late-time annihilations of the heavier scalar, in the context of the super-WIMP
mechanism.

\vfill

\vfill
\newpage
\tableofcontents
\end{titlepage}

\section{Introduction}
\setcounter{page}{3}
Dark matter (DM) comes as an interesting and challenging problem across astrophysics, cosmology and particle physics. Compelling evidence from accurate observations over many scales points to a non-baryonic and non-luminous matter component in the universe. Its nature is per se obscure, and dark matter could come in many forms such as primordial black holes, MACHOs, fundamental particles or a combination of them all. For an extensive and recent review on dark matter see e.g.~\cite{Bertone:2016nfn}. 

Over the last decades, one of the most studied options has been a weakly interacting massive particle (WIMP), that is assumed to share weak interactions with the visible sector, where we indeed mean of the order of magnitude of weak interactions in the Standard Model (SM). This option opens up a rich phenomenology, that has been tirelessly pursued by complementary experimental strategies like direct, indirect and collider searches. 
Nevertheless, the detection of a WIMP, and dark matter in general, has not been accomplished yet and many models are now put under strong tension with observations \cite{Arcadi:2017kky}. Perhaps this might be due to dark matter being feebly, far less than weakly, interacting with ordinary matter. 
In the end, it is the gravitational interaction of dark matter that
is mostly responsible for its successful role in forming the structures as observed in our universe. Other types of interactions with the visible sector may be absent or, at least, very much smaller than those assumed for a typical WIMP. 

Accordingly there has been renewed interest in a feebly interacting massive particle (FIMP) dark matter, whose production in the early universe is achieved through the \textit{freeze-in} mechanism \cite{McDonald:2001vt,Hall:2009bx} (see\cite{Baer:2014eja,Bernal:2017kxu} for reviews on the topic). In this framework, dark matter particles never reached thermal equilibrium due to their tiny coupling with the surrounding plasma, at variance with the central assumption of the \textit{freeze-out} mechanism. In the latter --- and extensively studied --- case, dark matter particles follow an equilibrium abundance when the temperature is larger than their mass, that it is also maintained when dark states enter a non-relativistic regime. Dark matter is mainly depleted by pair annihilations, which are very efficient up until $T / M \approx 1/20$. Around this temperature, dark matter particles decouple and this is the time its abundance is determined and frozen ever since. For freeze-in, the situation is rather the opposite. Dark matter particles never reach equilibrium due to a very small coupling with the plasma constituents. Dark matter particles are generated through the decays of a heavier accompanying state in the dark sector, $2 \to 1$ annihilations and $2 \to 2$ scatterings that may involve SM particles. Dark matter particles only appear in the final state of the relevant reactions, and their abundance builds up and increases over the thermal history. 

Let us remark that the relevant temperature range for  freeze-in is complementary with respect to that for freeze-out. This very fact holds for models with renormalizable interactions, where the dark matter production is dominated by $T \sim M$ \cite{McDonald:2001vt,Hall:2009bx} (also dubbed as infra-red freeze-in), as well as when non-renormalizable interactions are involved. In the latter case, usually referred to as  ultra-violet freeze-in \cite{Hall:2009bx,Yaguna:2011ei,Krauss:2013wfa,Elahi:2014fsa}, the production mechanism is sensitive to much higher temperatures, such as the reheating temperature which is set by reheating/end of inflation dynamics. This is again in contrast to the freeze-out mechanism, where thermal equilibrium erases all dependence on initial conditions.

To the best of our knowledge, the high-temperature range for renormalizable-interaction freeze-in produced dark matter has not been  addressed thoroughly. Only recently (i) the use of a Boltzmann distribution has been replaced by a more appropriate Fermi-Dirac/Bose-Einstein distribution for  the decaying particle~\cite{Belanger:2018ccd,Lebedev:2019ton,Bandyopadhyay:2020ufc}; (ii) the role of thermal masses has been investigated~\cite{Baker:2016xzo,Baker:2017zwx,Dvorkin:2019zdi,Darme:2019wpd}.  In the latter case, the effect of thermal masses has been explored in decay processes which would be forbidden at zero temperature and instead open up in a thermal plasma, and in association with phase transitions. 

In this work, we shall highlight the contribution to the dark matter production rate from multiple soft scatterings, that enhance the $1 \to 2$ decay process and more generally make effective $1\leftrightarrow 2$ processes
possible, in what is oftentimes called the Landau--Pomeranchuk--Migdal (LPM) effect, in analogy to its QED
counterpart \cite{Landau:1953gr,Landau:1953um,Migdal:1956tc}. We shall concentrate as well on the contribution to $2 \to 2$ processes in the ultra-relativistic regime, namely when the thermal scale $\pi T$ is larger than any other mass scale in the model (in-vacuum and thermal masses). Thermal masses, which are of order $gT$, play a role in both sets of processes, where $g$ here labels the parametrically more important couplings of
the equilibrated degrees of freedom. We build up on the developments carried out in the
production/equilibration rate of Majorana neutrinos
in the type-I seesaw leptogenesis framework \cite{Anisimov:2010gy,Besak:2012qm,Ghisoiu:2014mha,Ghiglieri:2016xye}. Also in this case, the Yukawa couplings are so small that the
Majorana fermions do not equilibrate.
Based on this analogy, we can immediately conclude that thermal masses, LPM
effect, as well as soft contribution to $2 \to 2$ scatterings 
might give a substantial contribution,
and we indeed find their numerical relevance being large. 

In order to illustrate these many effects,  we focus on a concrete model with a Majorana dark matter fermion accompanied by a heavier scalar particle, the latter sharing interactions with the Standard Model sector.  More precisely, we consider a simplified model often discussed in the literature and that has ties with the MSSM. In this model,  dark matter interacts with a SM quark via a colored scalar mediator.\footnote{Other realizations are of course possible, where the dark matter particle is a real scalar
or a vector boson \cite{Hisano:2011cs,DiFranzo:2013vra,An:2013xka,Garny:2015wea,Arina:2020udz}.} Such a model was already studied in the context of the freeze-in mechanism in refs.\cite{Hall:2009bx,Garny:2017rxs,Belanger:2018sti,Garny:2018icg,Garny:2018ali}. A similar model where there is a leptophilic interaction has been recently addressed in ref.\cite{Junius:2019dci}. On the phenomenological side, such models provide long-lived particle signatures due to a tiny coupling between  dark matter and the mediator \cite{Co:2015pka,Hessler:2016kwm,Junius:2019dci,Davoli:2017swj,Belanger:2018sti,Sirunyan:2018ldc,Aaboud:2019trc,Alimena:2019zri}. Moreover, the heavier mediator is typically responsible for additional dark matter production at much later stages in the thermal history via the so-called super-WIMP mechanism~\cite{Feng:2003xh,Feng:2003uy}. Here, the relic abundance of the mediator, as determined by pair annihilations and thermal freeze-out, is key to the extraction of the dark matter energy density. In this work, we shall include bound-state effects on the late-time annihilations of the colored mediator.

The structure of the paper is as follows. In Sec.~\ref{sec_model}, we introduce the dark matter model and its salient features. Then, we make the connection between the particle production rate as obtained from a spectral function with the standard Boltzmann approach in Sec.~\ref{sec_production_spectral}. Here, we compute the Born rate of the $1 \to 2$ process with and without thermal masses. Sec.~\ref{production_rate_UR} is devoted to the analysis and derivation of the production rate from multiple soft scatterings $1 +n \leftrightarrow 2 + n$ and $2 \to 2$ processes in the high-temperature regime, together with a phenomenological prescription to smoothly approach the relativistic/non-relativistic regime. Numerical results and comparisons of the relevant production rates, their thermal average and dark matter energy density, are collected in Sec.~\ref{numerical_results}. Finally, conclusions and outlook are offered in Sec.~\ref{concl_and_outlook}.

\section{Description of the simplified model}
\label{sec_model}
The simplified model that we consider consists of
a gauge singlet Majorana fermion $\chi$ and a scalar field
$\eta$. The latter 
is a singlet under SU(2)$_L$ but carries non-trivial QCD 
and hypercharge quantum numbers. In the MSSM framework, 
the Majorana fermion can be identified with a bino-like neutralino and the scalar 
with a right-handed stop or more generally any right-handed squark. However,
we do not fix couplings to their MSSM values and treat them as free parameters, and explore their impact on  dark matter production.

The Lagrangian for this extension of the SM can be 
expressed as \cite{Garny:2015wea} 
\begin{eqnarray}
 \mathcal{L} & = & 
 \mathcal{L}^{ }_{\hbox{\tiny SM}} + 
 \frac{1}{2} \, \bar{\chi} \left(  i \slashed{\partial} - M \right)  \chi 
 + (D^{ }_\mu \eta)^\dagger D^\mu \eta 
 - M_\eta^2\, \eta^\dagger \eta 
 - \lambda^{ }_2 (\eta^\dagger \eta)^2 
 \nonumber 
 \\
 & - & \lambda^{ }_3\, \eta^\dagger \eta\, \phi^\dagger \phi 
 - y\,  \eta^\dagger \bar{\chi} a_R q 
 - y^* \bar{q} a_L \chi\, \eta
 \;,
 \label{Lag_RT}
\end{eqnarray}
where $\phi$ is the SM Higgs doublet, $M_\eta$ the mass of the mediator and $M$ the mass of the DM particle, with $M_\eta> M$ so to ensure the fermion being the lightest, and stable, state of the dark sector. The Yukawa coupling between $\eta$ and $\chi$ is denoted by $y$, whereas $\lambda_2$ and $\lambda_3$  are the self-coupling of the coloured scalar and its portal coupling to the Higgs respectively. The coupling $\lambda_1$ is left for the SM Higgs self interaction and $a_R$ ($a_L$) is the right-handed (left-handed) projector. In this work, we set $\lambda_2=0$ in order to reduce the number  of free parameters of the simplified model.\footnote{%
It is worth noting that, to the order reached in this work, the only effect of a nonzero $\lambda_2$ would be an extra contribution
to the thermal mass of the $\eta$
boson for $T\gg M_\eta$, similar to the term proportional to $\lambda_3$
in Eq.~\eqref{asym_the_mass}. For 
$T\lesssim M_\eta$ this term would start to
become exponentially suppressed.}

In the following, we shall consider an unbroken SM phase; that amounts to restrict us to $T \simg 160$ GeV.\footnote{We also assume that the interactions between $\phi$ and $\eta$ do not affect the electroweak crossover. To this end, it suffices to take the colored scalar heavier than the light-stop window, where the scalar mass is of order of the top-quark mass \cite{Carena:1996wj,Delepine:1996vn,Cline:1996cr,Losada:1996ju,Laine:1996ms}.} In addition to the Yukawa interaction with the fermion $\chi$, the scalar $\eta$ shares interactions with QCD gluons, $B_\mu$ gauge boson of the U(1)$_Y$ SM gauge group, and with the Higgs doublet. Quarks are maintained in thermal equilibrium through QCD, SU(2)$_L$ and U(1)$_Y$ interactions, as well as SM Yukawa interactions; for heavy (top) quarks this interaction will also contribute to dark matter 
production at leading order. We denote with $Y_q$ the hypercharge of the SM quark ($Y_q=2/3, -1/3$ for up- and down-right handed quark respectively), whereas $h_q$ stands for the Yukawa coupling among a right-handed quark, a SU(2)$_L$ quark doublet and the Higgs field.

Let us now frame the model in the context of dark matter production in the early universe. 
As mentioned in the introduction, we assume the coupling $y$ between the dark matter $\chi$, the colored partner and the SM quarks to be very small ($y \simeq \mathcal{O}(10^{-7})$ or less \cite{Bernal:2017kxu}), such that the fermion $\chi$ never reaches thermal equilibrium. We further assume that a negligible population of dark matter is generated during reheating, whereas we take the reheating temperature $T \gg M_\eta$ so as to ensure that the colored scalars are in thermal equilibrium due to their gauge interactions. In this setting, there are two sources for the dark fermion production: the freeze-in mechanism, which is mostly efficient for $T \simg M_\eta$, and the super-WIMP mechanism \cite{Feng:2003xh,Feng:2003uy}, that accounts for the late decay of the frozen-out $\eta$ particles. In the latter case, the main ingredient is the energy density of the colored scalar at late times, or $T \ll M_\eta$, that is determined by pair annihilations. In this work, we include near-threshold effects, namely the Sommerfeld enhancement and bound-state formation, in order to extract the energy density of the $\eta$ particles (see Sec.~\ref{SW_contribution}). 

\section{Particle production rate from a spectral function}
\label{sec_production_spectral}
In this section we define our central quantity that we use throughout the paper, namely the particle production rate. We can express it in terms of a more fundamental object, which is the spectral function $\rho$ of the produced particle, here the Majorana fermion $\chi$. Most notably, and for practical computations, the spectral function can be related to the imaginary part of a retarded correlator at finite temperature $\textrm{Im} \Pi_R$, or an analytically continued Euclidean correlator $\textrm{Im} \Pi^E$ \cite{Asaka:2006rw,Bodeker:2015exa,Laine:2016hma}. 
The production rate for the dark fermion induces a change on the phase space distribution $f_\chi (t,\bm{k})$ as follows \cite{Bodeker:2015exa}
\begin{equation}
\label{diff_rate_eq}
\left( \frac{\partial}{\partial t} - H k_i \frac{\partial}{\partial k_i} \right) f_\chi (t,\bm{k})  = \Gamma(k)[n_\mathrm{F}(k^0)-f_\chi (t,\bm{k})],
\end{equation}
with
\begin{equation}
\Gamma(k) =
\frac{|y|^2}{k^0}\mathrm{Im}\Pi_R=
\frac{|y|^2}{2 k^0}  {\rm{Tr}} \left\lbrace  \slashed{\mathcal{K}} \, a_R \, \left[ \rho(\mathcal{K}) + \rho(-\mathcal{K}) \right]  \,  a_L \right\rbrace \, , 
\label{defgamma}
\end{equation}
where $k^0=\sqrt{k^2+M^2}$, $\mathcal{K}$ is the fermion four-momentum in Minkowski metric and $n_\mathrm{F}$ is the Fermi--Dirac 
distribution. Eq.~\eqref{diff_rate_eq} describes the production
and equilibration of the $\chi$ particles, parametrized
by a rate $\Gamma(k)$ that is given, to first order
in $y$ and to all orders in the SM+$\lambda_3$ couplings, by
Eq.~\eqref{defgamma}. The latter expresses the rate
as the imaginary part of the retarded self-energy of the $\chi$
fields. As demonstrated in \cite{Bodeker:2015exa}, this relation
shows that the so-called \emph{production rate}, defined by
taking the r.h.s. of Eq.~\eqref{diff_rate_eq} at negligible $f_\chi$,
as appropriate for a freeze-in production and as we shall indeed do later on, is identical to the \emph{equilibration rate}, 
valid when $f_\chi\sim n_\mathrm{F}$, to all orders in the SM+$\lambda_3$ couplings.

 We  furthermore consider an expanding background parametrized by the Hubble rate $H$ on the left-hand side of Eq.~(\ref{diff_rate_eq}). 
Moreover, the distribution $f_\chi$ in Eq.~\eqref{diff_rate_eq} represents the helicity-averaged distribution, therefore, a factor of 1/2 appears in the denominator of the right-hand side of Eq.~\eqref{defgamma}, where we included the contributions of the two helicities to the spectral function (in the end, since the fermionic spectral function is even in the absence of chemical potentials for the equilibrated species, i.e.~$\rho(-\mathcal{K})=\rho(\mathcal{K})$, the factors balance out). 

\begin{figure}[t!]
    \centering
    \includegraphics[scale=0.53]{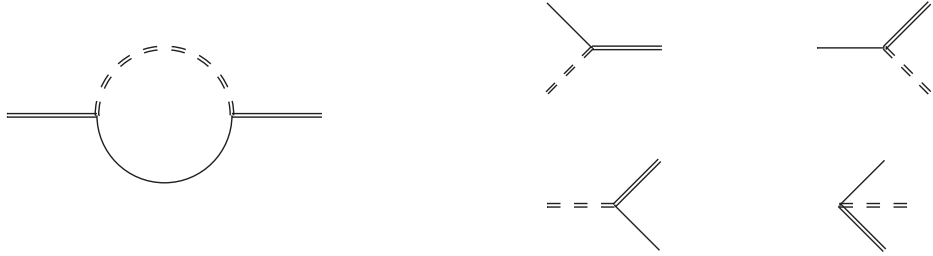}
    \caption{Left: one-loop self energy of the Majorana dark matter fermion, depicted with a solid-double line. The colored scalar is represented with a double-dashed line, whereas the solid line stands for the SM quark. Right: different channels originated by the cuts of the one-loop self-energy diagram according to the relation among the three particle masses; the produced particles are on the right of the vertex. The bottom-right process is never realized.}
    \label{fig:1to2_spectral}
\end{figure}

Non-trivial contributions to the spectral function can be computed by considering the Euclidean correlator of the operators coupling to the DM fermion through the interaction in (\ref{Lag_RT}), that corresponds
to the self-energy of the Majorana fermion. We show the diagram at leading order in Fig.~\ref{fig:1to2_spectral}. It reads
\begin{equation}
\Pi^E(K) \equiv\, {\rm{Tr}} \left\lbrace i \slashed{K} \left[  \int_X e^{i K \cdot X} a_R \langle (\eta^\dagger q) (X) (\bar{q} \eta)(0) \rangle a_L \right]  \right\rbrace  \, ,
\label{EC_1}
\end{equation}
where the Euclidean four-momentum $K$ is fermionic. Next, we carry out the field contractions and, by setting a vanishing in-vacuum mass for the quark, we obtain 
\begin{equation}
    \Pi^E(K) = N_c \int_{\bm{p}} T \sum_n \frac{-i\slashed{P}a_L}{p_n^2+E^2_q} \frac{1}{(p_n+k_n)^2+E_\eta^2} \, , 
    \label{EC_2}
\end{equation}
where we have defined the particle energies $E_q = |\bm{p}|=p$ and $E_\eta = \sqrt{(\bm{p+k})^2+M_\eta^2}$ for which we have inserted their free propagators. $N_c=3$ is the number of colours and $\int_{\bm p}\equiv\int d^3{\bm p}/(2\pi)^3$. One is left with the Matsubara sum and the Euclidean correlator in Eq.~(\ref{EC_2}) can be related to a spectral function through the methods exploited in refs.~\cite{Asaka:2006rw,Laine:2011pq}, and reviewed in detail in ref.~\cite{Laine:2016hma}. We find the spectral function of the dark matter fermion to be
\begin{eqnarray}
\rho (\mathcal{K}) = \frac{n_\mathrm{F}(k^0)^{-1}N_c}{2} \int_{\bm{p}_q, \bm{p}_\eta} \frac{\slashed{\mathcal{P}}_q a_L}{4 E_q E_\eta} (2 \pi)^4 \delta^4(\mathcal{P}_\eta - \mathcal{P}_q - \mathcal{K})  n_\mathrm{B}(E_\eta) (1-n_\mathrm{F}(E_q)) + \cdots
\end{eqnarray}
that accounts for the option $M_\eta > M$, namely the $1 \to 2$ decay process $\eta \to \chi q$. The dots correspond to complementary channels that are not realized in the model at leading order and the channel that never occurs (they are all displayed in Fig.~\ref{fig:1to2_spectral}).\footnote{If the quark had a vacuum mass $M_q$, the four channels in Fig.~\ref{fig:1to2_spectral} would correspond to (from left to right and top to bottom): $M>M_\eta +M_q$, $M_q > M + M_\eta$, $M_\eta > M + M_q$ and the last one never.} It is worth noticing that the thermal distributions for the colored scalar ($n_\mathrm{B}$ is the Bose--Einstein distribution) and SM quark appear as they do in (the gain term of) a Boltzmann equation. Upon defining the number density of dark matter particles as $n_\mathrm{DM}=2 \int_{\bm{k}} f_\chi(t,k)$, with the factor of $2$ accounting for the two helicity states, and upon
neglecting $f_\chi(t,k)\ll n_\mathrm{F}(k^0)$ on its r.h.s., we 
rewrite Eq.~(\ref{diff_rate_eq}) as follows 

\begin{eqnarray}
\dot{n}_\mathrm{DM} + 3H n_\mathrm{DM} &=& 2 |y|^2 \int_{\bm{k}} \frac{n_\mathrm{F}(k^0)}{k^0}  \textrm{Im} \Pi_R \\
&=& 2 |y|^2N_c (M_\eta^2 -M^2)  \int_{\bm{p}_\eta, \bm{p}_q, \bm{k}}  \frac{(2 \pi)^4 \delta^4(\mathcal{P}_\eta - \mathcal{P}_q - \mathcal{K})}{8 E_\eta E_q \, k^0}  n_\mathrm{B}(E_\eta) \left[ 1-n_\mathrm{F}(E_q)\right]
\nonumber\\
\label{number_density_Born}
\end{eqnarray}
Again we remark that Eq.~(\ref{number_density_Born}) is pretty much a Boltzmann equation for the production of dark fermions from colored scalar decays.

However, the clear advantage of the presented approach is its quantum-field theoretical nature, so that one can easily generalize it to higher order computations and thermal effects. 

\subsection{Born rate with vanishing thermal masses}
\label{sec_Born_vacuum}
In what follows, we call the production rate corresponding to the leading order process $\eta \to \chi q$ as the Born rate $\textrm{Im} \Pi_R^{\textrm{Born}}$. As a reference to single out the effects of thermal masses, we compute the leading order production rate in the case of vanishing thermal masses. This amounts to carrying out the phase-space integral in Eq.~(\ref{number_density_Born}). The mass scales relevant to our setting are the colored scalar mass $M_\eta$ and the dark matter mass $M$ (one could also trade one of the two with the mass splitting $\Delta M = M_\eta-M$). We use capital letters for in-vacuum masses. 

The computations share many similarities with the case of Majorana neutrinos in leptogenesis. For the model at hand we have two massive particles, the fermion $\chi$ and the scalar $\eta$, and the derivation of the Born rate agrees with that derived in leptogenesis when the Higgs mass is kept together with the Majorana neutrino mass \cite{Ghisoiu:2014ena,Ghiglieri:2016xye}. Upon integrating over $\bm{p}_\eta$ in Eq.~(\ref{number_density_Born}), we obtain 
\begin{eqnarray}
\textrm{Im} \Pi_R^{\textrm{Born}} &=&  \frac{N_c  (M_\eta^2 -M^2)}{8 n_\mathrm{F}(k^0)}\int \frac{d^3 \bm{p}_\eta}{(2 \pi)^3} \int \frac{d^3 \bm{p}_q}{(2 \pi)^3} \frac{(2 \pi)^4 \delta^4(\mathcal{P}_\eta - \mathcal{P}_q - \mathcal{K})}{ E_\eta E_q }  n_\mathrm{B}(E_\eta) (1-n_\mathrm{F}(E_q)) \nonumber 
\\
&=& 
\frac{N_c (M_\eta^2 -M^2)}{16\pi k} \int_{p_{\hbox{\tiny min}}}^{p_{\hbox{\tiny max}}} dp \left[ n_\mathrm{B}(p+k^0) + n_\mathrm{F}(p)\right] 
\nonumber 
\\
&=&
 \frac{N_c T (M_\eta^2 -M^2)}{16 \pi k} \left[ \ln \left( \frac{\sinh(\beta(k^0+p_{\hbox{\tiny max}})/2)}{\sinh(\beta(k^0+p_{\hbox{\tiny min}})/2)}\right) - \ln \left( \frac{\cosh(\beta p_{\hbox{\tiny max}} /2)}{\cosh(\beta p_{\hbox{\tiny min}}/2)}\right)\right] \, , 
 \label{vacuum_born}
 \end{eqnarray}
 where 
 \begin{equation}
     p_{\hbox{\tiny min}}=\frac{M^2_\eta - M^2 }{2(k^0+k)} \, , \quad p_{\hbox{\tiny max}}=\frac{M^2_\eta - M^2 }{2(k^0-k)} \, .
     \label{boundaries_vacuum_born}
 \end{equation}

\subsection{Born rate with finite thermal masses}
\label{sec_Born_full}
In order to address the freeze-in mechanism, we shall explore temperatures that range from $T \gg M_\eta$, the largest vacuum mass scale in the model, down to $T \siml M_\eta$ (even smaller temperatures are considered in the later stages of dark fermion production via the super-WIMP mechanism).  In the high-temperature limit, here $T \gg M_\eta$, the modification to the dispersion relation has to be taken into account and repeated interactions with the plasma constituents generate the so-called asymptotic masses. For the colored scalar and the SM quark, they read\footnote{As for the quark mass, we used the \textit{asymptotic mass} defined as $m_q^2 \equiv 2 m_F^2$ according to ref.\cite{Laine:2016hma}.} 
\begin{eqnarray}
m^2_\eta= \left( \frac{g_3^2C_F+Y_q^2g_1^2}{4} + \frac{\lambda_3}{6}  \right)  T^2 \, , \quad m_q^2= \frac{ T^2}{4}(g_3^2C_F + Y_q^2g_1^2 + |h_q|^2) \, ,
\label{asym_the_mass}
\end{eqnarray}
where we note in passing that the gauge contribution is the same for both the scalar and fermion; there, $C_F=(N_c^2-1)/(2N_c)$ is the quadratic Casimir of the fundamental 
representation. The thermal mass for the DM is negligible since it is proportional to $|y|^2 \ll g_3^2, g_1^2, \lambda_3,|h_t|^2$. These  are the couplings
that we consider parametrically of the same order and that we label
collectively as $g=(g_1,g_3,\sqrt{\lambda_3},h_t)$; $g_1$ and $g_3$ are the U(1)$_Y$
and SU(3) gauge couplings. The thermal asymptotic masses in Eq.~(\ref{asym_the_mass}) come from the $p^0\sim p\sim \pi T$,
$p^0-p\ll \pi T$ limit of the thermal self-energies for these 
particles. This limit generally agrees with the $p^0\to p$
limit of the Hard Thermal Loop (HTL) self-energies.
The latter arise in the limit $p^0,p\sim gT\ll \pi T$ and exploit
the scale separation between the hard momentum scale in the loop, which is $\pi T$, and the soft external momenta scaling as $g T$. A consistent EFT resumming these (and higher-point) amplitudes, going under the name of HTL resummmation \cite{Braaten:1989mz,Frenkel:1989br,Taylor:1990ia,Frenkel:1991ts,Braaten:1991gm},
will play an important role in the remainder of this paper.
Within the HTL theory, 
screening thermal masses, also known as Debye masses, are generated for the U(1)$_Y$ and SU(3) gauge boson, and they read
\begin{equation}
    m_{\hbox{\tiny B}}^2= \left( \frac{n_{\hbox{\tiny S}}}{6} + \frac{5n_{\hbox{\tiny G}}}{9} + \frac{Y_q^2 N_c}{3}\right) g^2_1 T^2 \, , \quad m^2_g=\left( \frac{N_c}{3} + \frac{n_{\hbox{\tiny G}}}{3} + \frac{1}{6} \right) g^2_3 T^2 \, ,
    \label{HTL_gauge_bosons}
\end{equation}
where we have taken into account the contribution due to $\eta$ in the gauge boson thermal self-energies (last term in each of the masses in Eq.~(\ref{HTL_gauge_bosons})), $n_{\hbox{\tiny S}} =1$ is the number of Higgs doublets and $n_{\hbox{\tiny G}}=3$ the number of fermionic generations.
\subsubsection{Asymptotic scalar mass at non-vanishing vacuum mass}
The asymptotic thermal mass of the scalar is not a good approximation when the vacuum mass $M_\eta$ is no longer negligible with respect to the scale $gT$. As soon as 
$M_\eta\gtrsim  T$, one must furthermore include it in the determination of the thermal self-energy of the 
scalar, which in turn gives the thermal contribution to the mass (asymptotic mass) $m_\eta$.
As Eq.~\eqref{asym_the_mass} shows, the latter is given by a gauge contribution and a
portal coupling contribution. The gauge contribution contains a tadpole, from the ``seagull'' gauge-gauge-scalar-scalar vertex, and
an external momentum-dependent term from the two scalar-gauge-scalar vertices. As the tadpole features  a gauge boson loop, it will not
be affected by the scalar mass, while the momentum-dependent diagram will be affected, as one of the
two propagators is the scalar's. Finally, the tadpole induced by the portal
coupling features a Higgs scalar in the loop and is thus independent of $M_\eta$.

Care must be taken regarding the gauge dependence of the gauge contribution. At vanishing mass, the scalar self-energy
in the HTL limit is gauge invariant and momentum independent. It stays gauge-invariant
outside of the HTL limit when the external momentum $\mathcal{P}=(p^0,\bf{p})$ obeys $p^0\sim p\sim \pi T$, $p^0-p\ll \pi T$, that is, on the light-cone limit, which
defines the asymptotic mass (thus gauge invariant, as it is a pole mass). Similarly, at finite mass one finds that the
self-energy in the limit $\mathcal{P}^2\to M_\eta^2$ is gauge invariant, which we have explicitly checked 
by performing the calculation in Feynman and Coulomb gauge.
\begin{figure}[t!]
    \centering
    \includegraphics[width=10cm]{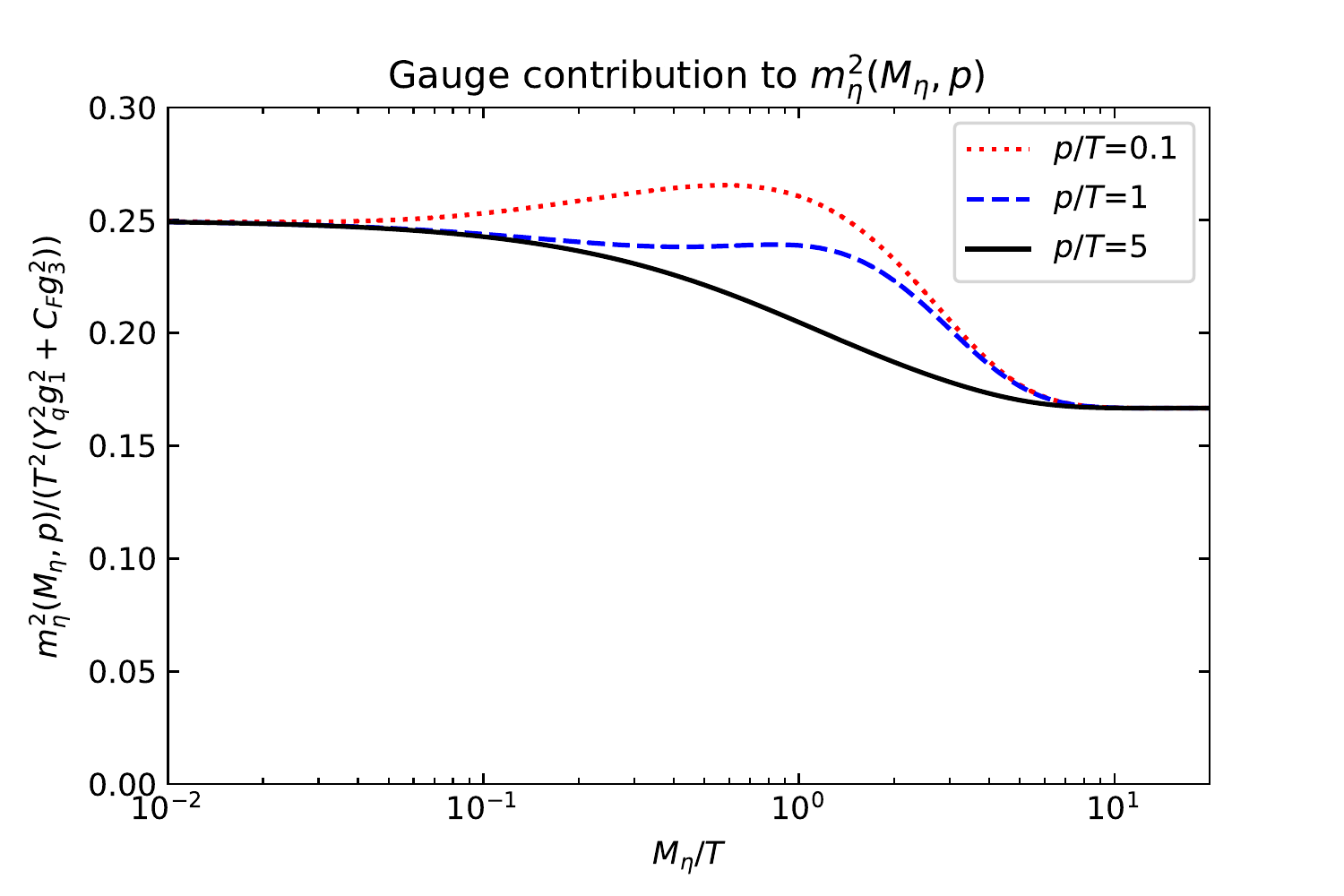}
    \caption{The gauge contribution to $m^2_\eta(M_\eta,p)$ in Eq.~\eqref{scalartadpole}, as a function of $M_\eta$ and $p$.
    The quartic scalar contribution is momentum-independent.}
    \label{fig_asym_mass}
\end{figure}
We can thus define
\begin{equation}
    m^2_\eta(M_\eta,p) = \mathrm{Re}\;\Pi^\eta_R\left(p^0=\sqrt{p^2+M_\eta^2},p\right).
    \label{defmeta}
\end{equation}
 We find
 \begin{align}
     m^2_\eta(M_\eta,p) =\; & \frac{Y_q^2g_1^2+C_F g_3^2}{2\pi^2}\int_0^\infty dq \bigg\{\frac{n_\mathrm{B}(E_\eta(q))}{E_\eta(q)}\bigg[q^2+
     \frac{M_\eta^2 q}{2p}\ln\left(\frac{(p+q)^2}{(p-q)^2}\right)\bigg]+2\, q\, n_\mathrm{B}(q)\bigg\} \nonumber \\
     &+\frac{\lambda_3}{\pi^2}\int_0^\infty dq\,q\,n_\mathrm{B}(q),
     \label{scalartadpole}
 \end{align}
where the first line is the gauge contribution and the second the portal one. We see that, at vanishing mass 
$E_\eta(q)\to q$, Eq.~\eqref{asym_the_mass} is recovered, whereas in the non-relativistic limit $M_\eta\gg T$ we find
\begin{equation}
\label{nrmass}
    m^2_\eta(M_\eta\gg T,p ) = (Y_q^2g_1^2+C_F g_3^2+\lambda_3)\frac{T^2}{6}.
\end{equation}
Between these two limiting cases, $m_\eta$ is also $p$-dependent, though in practice the dependence is quite mild
and the transition between the two regimes rather smooth, as shown in Fig.~\ref{fig_asym_mass}.

In the non-relativistic limit Eq.~\eqref{nrmass} is not a good approximation, as
the so-called Salpeter term, which is negative, is of comparable size, if not larger. It arises from the contribution of screened soft gluons at the
scale $gT$ and it yields \cite{Biondini:2018pwp}
\begin{equation}
    m^2_\eta(M_\eta\gg T,p ) = (Y_q^2g_1^2+C_F g_3^2+\lambda_3)\frac{T^2}{6}-\frac{M_\eta}{4\pi}(g_1^2Y_q^2 m_{\mathrm{B}}+g_3^2 C_F m_{g}). 
\end{equation}
As the dispersion relation is determined by the zeros of the denominator of the propagator, which features the sum of the in-vacuum and thermal mass squared, $\mathcal{M}_\eta^2 \equiv M_\eta^2 + m_\eta^2$,
it follows that, in the non-relativistic limit
\begin{eqnarray}
\mathcal{M}_\eta \approx M_\eta+ (Y_q^2g_1^2+C_F g_3^2+\lambda_3)\frac{T^2}{12 M_\eta}-\frac{1}{8\pi}(g_1^2Y_q m_{\mathrm{B}}+g_3^2 C_F m_{g}),
\end{eqnarray}
whose $\lambda_3$ and $g_3$ contributions agree with \cite{Biondini:2018pwp}. Let us 
remark that, even though the thermal contribution to $\mathcal{M}_\eta$
is much smaller than $M_\eta$ in the non-relativistic regime $T\ll M_\eta$,
the Born rate depends on the combination $\mathcal{M}_\eta^2-M^2-m_q^2$, as we will
show momentarily. Thus, for $M_\eta-M\ll M_\eta$, the thermal contribution to $\mathcal{M}_\eta$ is not negligible even in the non-relativistic regime.

\subsubsection{Treatment of the quark thermal mass}
Since we look at temperatures larger than the electroweak crossover, the thermal mass is the only contribution for the SM quark, as its Higgs-mechanism vacuum mass vanishes there. There are some subtle differences between the fermion thermal mass and the scalar thermal mass. 
At vanishing vacuum mass, the scalar thermal mass in Eq.~\eqref{asym_the_mass} is momentum-independent, so that it is the same both for $p\to 0$ and for $p\gtrsim \pi T$. This is not the case for
the quark thermal mass, which is momentum-independent only for $p\gg gT$: as soon as $p$ becomes
soft, the full momentum dependence of the quark HTL dispersion
relation becomes necessary. However, when deriving the
Born rates, the regions where $p\sim gT$ represent subleading slices of the phase space, so that
their precise form is not necessary.\footnote{Only in cases
where the process happens very close to threshold, e.g. $\mathcal{M}_\eta\gtrsim m_q+M$, would they become more relevant.} In what follows, we will then use the UV limit of the quark
HTL propagator, which gives the correct dispersion relation at $p\gg gT$. It reads 
\begin{equation}
    S_R(p^0,p\gg gT)=\frac{i h^+_{\bf p}}{p^0-p-m_q^2/(2p)+i\epsilon}+\frac{i h^-_{\bf p}}{p^0+p+m_q^2/(2p)+i\epsilon},
    \label{asymhtl}
\end{equation}
where $h^\pm_{\bf p}=(\gamma^0\mp\bm{\gamma}\cdot\bm{p})/2$.
For $p\gg gT$, where Eq.~\eqref{asymhtl} is valid, the following form also holds
\begin{equation}
    S_R(p^0,p\gg gT)=\frac{i h^+_{\bf p}}{p^0-\sqrt{p^2+m_q^2}+i\epsilon}+\frac{i h^-_{\bf p}}{p^0+\sqrt{p^2+m_q^2}+i\epsilon},
    \label{asymhtl2}
\end{equation}
and it is the one that we use to determine the Born rate.

A crucial difference with respect to the Born rate computed with vanishing thermal masses is that two options can be realized: $\mathcal{M}_\eta > M + m_q$ or $m_q > \mathcal{M}_\eta + M$ depending on the temperatures and values of the couplings. In other words, the dark fermion can also be produced in the decay of the quark at high temperatures. 

In the $\eta$ decay case, we find 
\begin{equation}
\label{fullborneta}
    {\rm{Im}}\Pi_{\hbox{\tiny R},\eta \to \chi q}^{\mathrm{Born}}=\frac{N_c}{16\pi k}\int_{p_\mathrm{min}}^{p_\mathrm{max}} d p [\mathcal{M}_\eta^2-M^2-m_q^2-2 k^0 (E_p-p)]  [n_\mathrm{B}(k^0+E_p)+n_\mathrm{F}(E_p)],
\end{equation}
with $E_p=\sqrt{p^2+m_q^2}$ and the integration boundaries are
\begin{equation}
    p_\mathrm{min,\,max}=\frac{\mathcal{M}_\eta^2-M^2-m_q^2}{2M^2}\left|k^0\sqrt{1-\frac{4M^2 m_q^2}{(\mathcal{M}_\eta^2-M^2-m_q^2)^2}}\mp k\right|.
    \label{boundaries_thermal_born}
\end{equation}
We note that, had we used a vacuum mass term for $m_q$, we would have
\begin{equation}
    [\mathcal{M}_\eta^2-M^2-m_q^2-2 k^0 (E_p-p)] \to p\frac{\mathcal{M}_\eta^2-M^2-m_q^2}{E_p}.
\end{equation}
The two expression agree very well as long as $p\gg g T$ ($E_p-p\sim g^2 T$) and $\mathcal{M}_\eta^2-M^2\gg m_q^2$. When close to the threshold ($\mathcal{M}_\eta^2-M^2\sim m_q^2$) the approximation we have used
reflects the full HTL dynamics much better than the vacuum mass term would.
As a sanity check, we can take the $m_q\to 0$ and $\mathcal{M}_\eta \to M_\eta$ limits and simply recover from Eq.~\eqref{fullborneta} and (\ref{boundaries_thermal_born}) the former result in Eq.~(\ref{vacuum_born}) with the integration boundaries as in Eq.~(\ref{boundaries_vacuum_born}).
In the $q$ decay case we find instead
\begin{equation}
\label{fullbornq}
    {\rm{Im}}\Pi_{\hbox{\tiny R},q \to \eta \chi}^{\mathrm{Born}}=\frac{N_c}{16\pi k}\int_{p_\mathrm{min}}^{p_\mathrm{max}} d p [m_q^2-\mathcal{M}_\eta^2+M^2-2 k^0 (E_p-p)]  [n_\mathrm{B}(E_p-k^0)+n_\mathrm{F}(E_p)],
\end{equation}
where now
\begin{equation}
    p_\mathrm{min,\,max}=\frac{m_q^2+M^2-\mathcal{M}_\eta^2}{2M^2}\left|k^0\sqrt{1-\frac{4M^2 m_q^2}{(\mathcal{M}_\eta^2-M^2-m_q^2)^2}}\mp k\right|.
\end{equation}

\section{Production rate in the ultra-relativistic regime}
\label{production_rate_UR}
In the freeze-in production mechanism, the accumulation of the dark matter particles takes place over a broad range of temperatures. Typically, any process that generates a dark particle in the final state contributes to the dark matter energy density. In this model, the dark fermion $\chi$ is produced both via decays and scattering processes, that have to be studied in the temperature regime where SM particles and the accompanying $\eta$ scalar are in thermal equilibrium. Hence, we study the evolution of the abundances 
starting from temperatures $T \gg M_\eta$; there, all the involved particles are ultra-relativistic. In this setting, all the particles are essentially seen as  massless with respect to the hard scale $\pi T$ typical of the thermal motion, and hence the angle between the initial state momenta or between
the final state momenta can be small. This defines the collinear kinematics as a distinctive feature of the high-temperature dynamics. 

In the following Secs.~\ref{LPM_rate} and \ref{2to2_rate} we study two class of processes: 
effective $1\leftrightarrow 2$ processes, induced by multiple soft scatterings, and $2 \leftrightarrow 2$ scatterings, whose correct derivation at leading order need both HTL and LPM resummation. 

\subsection{LPM resummation for light-cone kinematics}
\label{LPM_rate}
\begin{figure}[t!]
    \centering
    \includegraphics[scale=0.49]{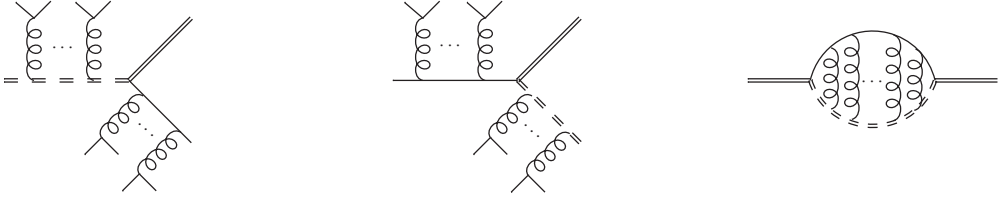}
    \caption{Leftmost and middle diagram: processes where the QCD gluons induce soft scatterings with plasma constituents (coming from the HTL resummed gluon propagator). Both 
    coloured/charged particles ($\eta$, $q$) can undergo multiple scatterings. The same processes occur with the U(1)$_Y$ gauge boson. Rightmost diagram: self-energy of the Majorana fermion with an arbitrary number of soft gauge boson
ladder rungs.}
    \label{fig:1to2_soft}
\end{figure}
At high temperatures all external momenta can be taken as hard, $p \sim \pi T$, whereas the thermal masses of the particles define a soft scale of order $gT$. The thermal mass of the dark fermions is $\sim y T$, which is way smaller than the thermal masses for the colored scalar $\eta$ and the quark.
In the regime $T \gg M_\eta,M$, we can treat the vacuum masses $M$
and $M_\eta$ as  soft scales. All the particles involved in the reactions are then close to the light cone because $\mathcal{P}^2 \sim (gT)^2$, where $\mathcal{P}$ is a Minkowskian four-momentum.

Let us now consider multiple scatterings with plasma constituents, as shown in the leftmost and middle diagram in Fig.~\ref{fig:1to2_soft}. Though these processes seem to be of higher order, due to the many additional vertices, \textit{they are not}. The gauge bosons that are exchanged
with the thermal constituents of the medium are soft, $q^0\sim q\sim g T$: they are thus 
described by the HTL expression that gives the correct screening mass when the external momentum is spacelike (see Eq.~(\ref{HTL_gauge_bosons})). These soft exchanges make
the intermediate virtual $\eta$ bosons and quarks almost on shell, with a lifetime (or \emph{formation time}) of
order $1/g^2T$, which is long and parametrically of the same order of the
soft-gauge-boson-mediated scattering rate (see e.g. \cite{Ghiglieri:2020dpq}). Hence, many of these soft scatterings
can take place during the formation time of the outgoing pair, so that their
quantum-mechanical interference, the LPM effect, has to be accounted for in a procedure
called LPM resummation. Under this procedure, we speak of \emph{effective} $1\leftrightarrow 2$
processes to describe the $1 + n \leftrightarrow 2 +n$ processes being consistently
accounted for at leading order. In the context of freeze-in production rates in the early
universe, LPM resummation was introduced for right-handed Majorana neutrinos in \cite{Anisimov:2010gy,Besak:2012qm,Ghisoiu:2014ena} and \cite{Ghiglieri:2016xye} in
the symmetric and broken electroweak phases, respectively.

In order to make the connection between the effective $1\leftrightarrow 2$ processes  and the imaginary part of a retarded correlator $\textrm{Im}\Pi^{\hbox{\tiny LPM}}_R$ (or the corresponding spectral function) we can consider the self-energy of the fermion $\chi$ with an arbitrary number of soft-gluon rungs connecting the scalar and the quark, as shown in Fig.~\ref{fig:1to2_soft} (rightmost diagram). These soft gluons also give rise to self-energy
contributions for the $\eta$ and the quark, which have to be resummed in their propagators 
and are not shown graphically.

Having delineated the physics of the $1 + n \leftrightarrow 2 + n$ processes, we can now pass to the main technical ingredients one needs to compute $\textrm{Im}\Pi^{\hbox{\tiny LPM}}_R$. We borrow the notation and computational setting from refs.~\cite{Anisimov:2010gy,Besak:2012qm,Ghisoiu:2014ena,Ghiglieri:2016xye}, where the production rate of a Majorana fermion involves the Higgs doublet and a lepton in the context of leptogenesis. First, we define an effective Hamiltonian 
\begin{equation}
    \hat{H} \equiv -\frac{M^2}{2k_0} + \frac{m_q^2-\nabla^2_\perp}{2 E_q} +  \frac{\mathcal{M}_\eta^2-\nabla^2_\perp}{2 E_\eta} +i\Gamma(y) \, , \quad y \equiv |\bm{y}_\perp| \, ,
\end{equation}
where $\nabla_\perp$ is a two-dimensional gradient that operates in two directions orthogonal to $\bm{k}$, the momentum of the dark fermion. 
 $\Gamma(y)$ encodes soft gauge scatterings; it can be expressed as
\begin{equation}
    \Gamma (y) =\frac{T}{2 \pi} g_1^2 Y_q^2  \left[  \ln \left( \frac{m_{\hbox{\tiny B}}y}{2}\right) + \gamma_E +K_0(m_{\hbox{\tiny B}}y) \right] 
    + \frac{T}{2 \pi} g_3^2 C_F  \left[  \ln \left( \frac{m_g  y}{2}\right) + \gamma_E +K_0(m_g y)\right]\, ,
\end{equation}
where the first term originates from U(1)$_Y$ gauge boson with Debye mass $m_\mathrm{B}$, and the second term from QCD gluons, with Debye mass $m_g$; $K_0$ is a modified Bessel function. 

Next, the effective Hamiltonian enters the inhomogeneous equations for the functions $g(\bm{y})$ and $\bm{f}(\bm{y})$ that are necessary to compute the LPM effect
\begin{equation}
    (\hat{H}+i0^+)g(\bm{y})=\delta^{(2)}(\bm{y}) \, , \quad   (\hat{H}+i0^+) \bm{f}(\bm{y})=-\nabla_{\perp} \delta^{(2)} (\bm{y}) \, .
    \label{LPM_diff_fg}
\end{equation}
In terms of  $g(\bm{y})$ and $\bm{f}(\bm{y})$, the production rate can be expressed as follows\footnote{When setting $N_c=2$ for the SU(2)$_L$ Standard Model gauge group, Eq.~\eqref{lpm_eq} agrees with Eq.~(4.5) of ref.\cite{Ghisoiu:2014ena}, that in turn conforms to the original derivation in ref.\cite{Anisimov:2010gy}.}
\begin{eqnarray}
{\rm{Im}}\Pi_{\hbox{\tiny R}}^{\hbox{\tiny LPM}}&=& -\frac{N_c}{8 \pi} \int_{-\infty}^{+\infty} d E_q  \int_{-\infty}^{+\infty} d E_\eta \, \delta(k_0-E_q-E_\eta) [1-n_\mathrm{F}(E_q)+n_\mathrm{B}(E_\eta)] \nonumber 
\\
&& \frac{k_0}{E_\eta} \lim_{\bm{y}\to0} \left\lbrace  \frac{M^2}{k_0^2} {\rm{Im}} [g(\bm{y})] + \frac{1}{E_q^2} {\rm{Im}} [\nabla_\perp \cdot \bm{f}(\bm{y})]\right\rbrace \, . 
\label{lpm_eq}
\end{eqnarray} 
At this stage, we can perform a non-trivial check of the LPM rate in Eq.~(\ref{lpm_eq}), namely taking the limit of vanishing soft scatterings to recover a \textit{collinear} Born rate. A detailed derivation of it, in the case of QCD photo-production, is given in ref.~\cite{Ghiglieri:2014kma}, and for $\mathcal{M}_\eta>M+m_q$, the LPM rate (\ref{lpm_eq}) reduces to
\begin{equation}
\label{collborn}
    {\rm{Im}}\Pi_{\hbox{\tiny R}}^{\hbox{\tiny LPM}\,\mathrm{Born}}=\frac{N_c}{16\pi k_0}\int_{E_{\mathrm{min}}}^{E_\mathrm{max}} d E_q \frac{E_q(\mathcal{M}_\eta^2-M^2-m_q^2)-k_0 m_q^2}{E_q} [n_\mathrm{B}(k_0+E_q)+n_\mathrm{F}(E_q)],
\end{equation}
with
\begin{equation}
    E_\mathrm{min},E_\mathrm{max}=\frac{k_0}{2M^2}\bigg(\mathcal{M}_\eta^2-M^2-m_q^2\mp\sqrt{(\mathcal{M}_\eta^2-m_q^2)^2+M^4-2 M^2(\mathcal{M}_\eta^2+m_q^2)}\bigg).
\end{equation}
In the $m_q\to 0$ and  $\mathcal{M}_\eta \to M_\eta$ limit, we obtain
\begin{equation}
    {\rm{Im}}\Pi_{\hbox{\tiny R}}^{\hbox{\tiny LPM}\,\mathrm{Born}}\bigg\vert_{m_q=0}=\frac{N_cT(M_\eta^2-M^2)}{16\pi k}\ln\frac{2 n_\mathrm{B}(k_0)(e^{k_0 M_\eta^2/(M^2T)}-1)}{e^{k_0 M_\eta^2/(M^2T)-k_0/T}+1},
\end{equation}
which agrees with the collinear limit of Eq.~\eqref{vacuum_born}, i.e. taking $k^0-k\ll k^0+k$.

It is worth remarking that the integration region in Eq.~\eqref{lpm_eq} accounts both for
effective $1\to 2$ processes and for effective $2\to 1$ processes. Once $E_\eta$ is fixed
to $k^0-E_q$ by the $\delta$ function, there are three distinct regions\footnote{In this list
we do not distinguish between particle and antiparticle states. Strictly speaking we should have e.g. $\bar{\eta},q\to \chi$.}
\begin{enumerate}
    \item  $k^0>E_q>0$: this corresponds to the effective $2\to 1$ process $\eta,q\to \chi$,
    \item \label{etadecay} $E_q<0$: this corresponds to the effective $1\to 2$ process $\eta\to q \chi$. Eq.~\eqref{collborn} is the $n=0$ limit (no scatterings) thereof,
    \item $E_q>k^0$: this corresponds to the effective $1\to 2$ process $q\to \eta \chi$.
\end{enumerate}
Without accounting for soft scatterings, at most one of these three scenarios is realized at a 
time, e.g. scenario~\ref{etadecay} for $\mathcal{M}_\eta>m_q+M$. The inclusion of soft
scatterings makes all three scenarios kinematically
allowed at the same time.

\subsubsection{Numerical strategy and subtraction of the Born limit}
In the interest of practicality, it is helpful to reduce the solution of the inhomogeneous equations (\ref{LPM_diff_fg}) to the regular solutions of the corresponding homogeneous equations with specific angular quantum numbers \cite{Strassler:1990nw,Ghisoiu:2014mha}, here $u^r_\ell$. Upon the definition of a dimensionless distance $\rho = y m_g$, the homogeneous equation reads
\begin{equation}
    \left[ -\frac{d^2}{d \rho^2} + \frac{\ell^2 -1/4}{\rho^2} + \frac{M_{\textrm{eff}}^2}{m^2_g} +  i \frac{2 \omega (k_0-\omega)}{k_0 m^2_g} \Gamma\left( \frac{\rho}{m_g^2} \right)  \right]u^{r}_{\ell}(\rho)=0 \, ,
\end{equation}
where the effective mass is
\begin{equation}
     M_{\textrm{eff}}^2 \equiv  -\frac{M^2 \omega (k_0-\omega)}{k_0^2} +\frac{m_q^2 (k_0-\omega)}{k_0} + \frac{\mathcal{M}_\eta^2 \, \omega}{k_0} \, .
\end{equation}
The solution for the scalar and vector functions $g(y)$ and $\bm{f}(\bm{y})$ corresponds to the angular quantum number $\ell=0$ and $\ell=1$ respectively. The LPM rate becomes 
\begin{eqnarray}
    {\rm{Im}}\Pi_{\hbox{\tiny R}}^{\hbox{\tiny LPM}} &=& -\frac{N_c}{2 \pi^2} \int_{-\infty}^{+\infty} d \omega [1-n_\mathrm{F}(\omega)+n_\mathrm{B}(k_0+\omega)] 
    \nonumber \\
    &&\left\lbrace \frac{M^2 \omega}{4 k_0^2} \int_0^{\infty} d\rho \, \textrm{Im} \left[ \frac{1}{[u^r_{0}(\rho)]^2}\right] + \frac{m^2_g}{\omega} \int_0^{\infty} d\rho \, \textrm{Im} \left[ \frac{1}{[u^r_1(\rho)]^2}\right] \right\rbrace \, .
    \label{LPM_numerics}
\end{eqnarray}
\begin{figure}[t!]
    \centering
    \includegraphics{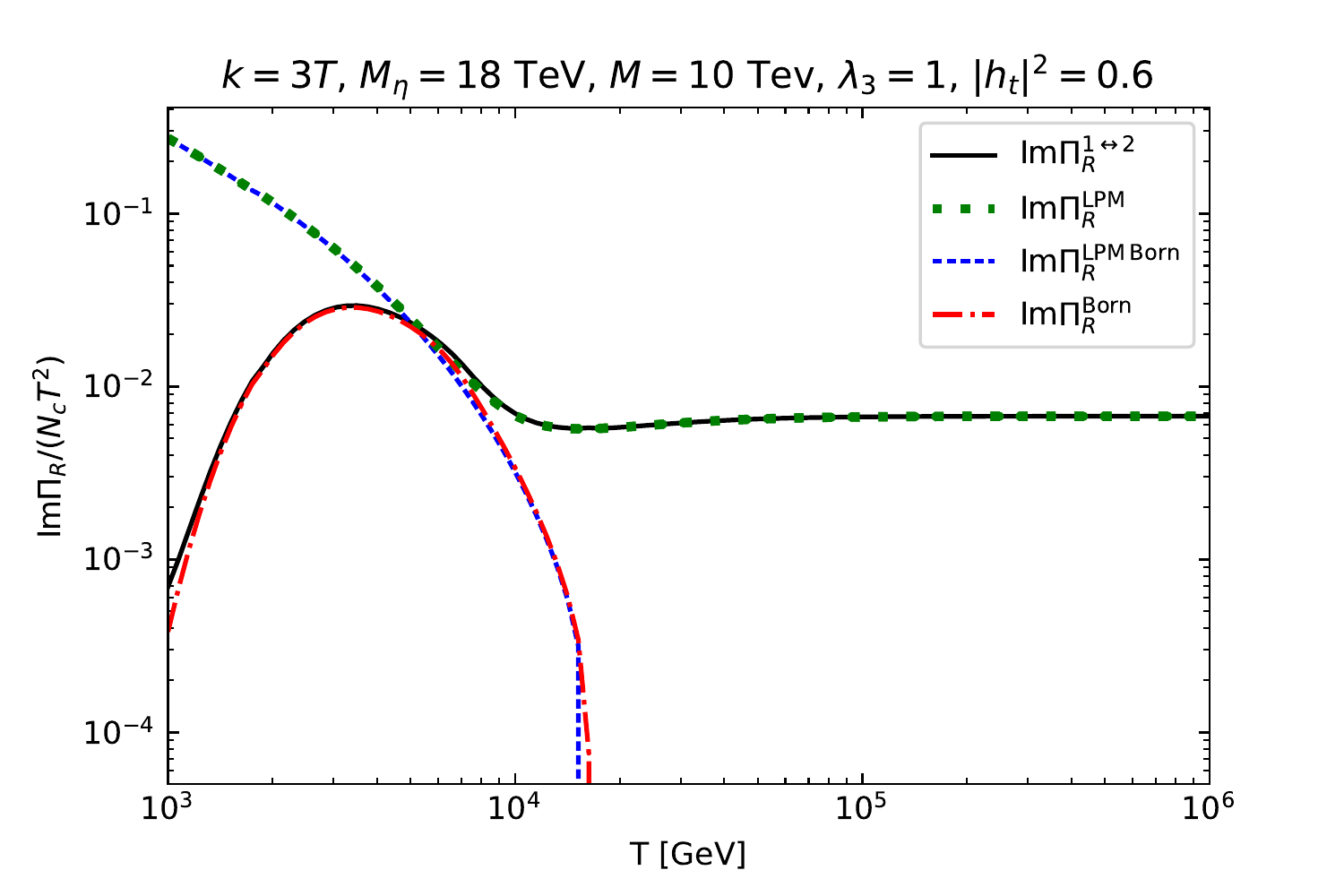}
    \caption{In this figure, the strong coupling is fixed at $g_3^2=0.85$, and the effect of the $U(1)$ gauge coupling is neglected, i.e. $g_1=0$.
    The values
    for the masses listed above the plot are the zero-temperature masses. $m_q^2$ is taken from 
    Eq.~\eqref{asym_the_mass}. The total mass squared of the $\eta$
    is given by the sum of the vacuum $M_\eta^2$ with Eq.~\eqref{scalartadpole} and, for $M_\eta>10 T$, the negative 
    Salpeter contribution as well. In the case of Eqs.~\eqref{lpm_eq}
    and Eq.~\eqref{collborn}, Eq.~\eqref{scalartadpole} is evaluated at the
    momentum of the $\eta$ boson, while for Eq.~\eqref{fullborneta}
    we keep the momentum fixed to a typical thermal momentum of $3T$
    (as Fig.~\ref{fig_asym_mass} shows, the momentum dependence is small).
    }
    \label{fig_subtr_ht}
\end{figure}

We are now in the position to compare the $1 \to 2$ Born rate in Eq.~(\ref{fullborneta}) 
with the $1+n \leftrightarrow 2 + n$ LPM rate and its Born (collinear) limit. Our aim is to 
test the various prescriptions
for the Born term and to define a recipe to make the LPM results
more accurate once (some of) the masses are no longer much smaller than the 
temperature, thus causing the failure of the collinear approximation. The 
state-of-the-art analyses brought forward in \cite{Ghisoiu:2014ena} for
right-handed neutrino production and in \cite{Ghisoiu:2014mha,Ghiglieri:2014kma} for dilepton production are not
directly applicable to this case. Indeed, they require knowledge of the
NLO contribution to $\mathrm{Im}\Pi_R$ in the relativistic regime, $M\sim T$,
for a massive external state (the $\chi$) \textit{and} for a massive
internal one (the $\eta$), which is significantly more intricate than
in the case of dileptons or sterile neutrinos. There only the external
state is massive, but the relativistic-regime calculations are nonetheless
highly nontrivial \cite{Laine:2013vpa,Laine:2013lka,Laine:2013vma,Jackson:2019mop}.
We shall thus content ourselves with a simpler prescription ensuring
a reasonable $1\leftrightarrow2$ rate at all values of $M/T$.
To this end, we shall take the LPM result, obtained by solving 
Eq.~\eqref{LPM_numerics}, subtract from it its Born term, Eq.~\eqref{collborn},
and replace it with the general-kinematics Born term in  Eq.~(\ref{fullborneta}),
which also accounts for the quark thermal mass, so that
\begin{equation}
    \label{sub_pres}
    \mathrm{Im}\Pi_R^{1\leftrightarrow 2}=\mathrm{Im}\Pi_R^\mathrm{LPM}-
    \mathrm{Im}\Pi_R^\mathrm{LPM\;Born}+\mathrm{Im}\Pi_R^\mathrm{Born}.
\end{equation}
\begin{figure}[t!]
    \centering
    \includegraphics{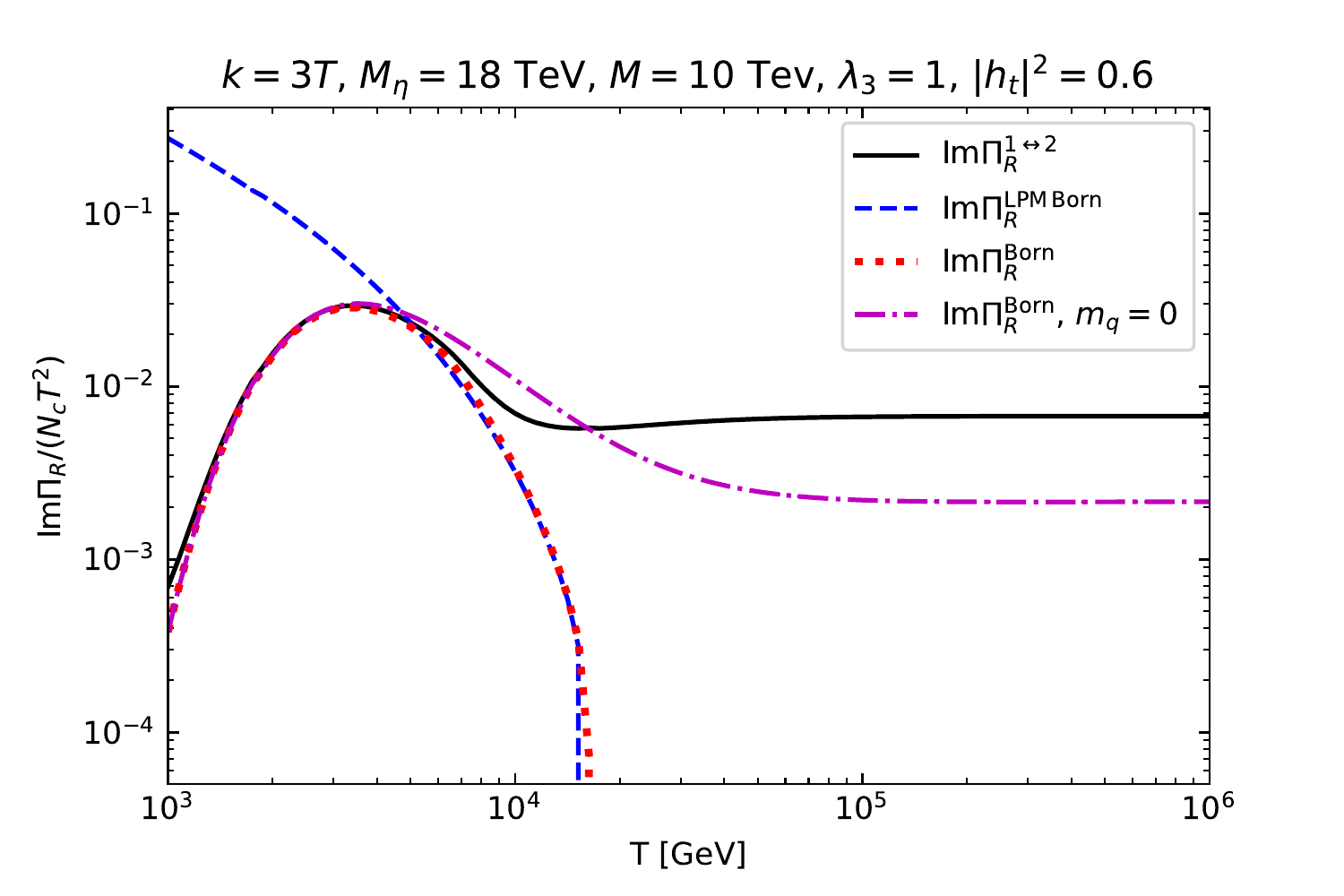}
    \caption{The effect of the different prescriptions for the Born
    term, shown alongside the curve from Eq.~\eqref{sub_pres}. Here, the Born rate is also taken with a vanishing thermal mass for the quark. Parameters
    are fixed as in Fig.~\ref{fig_subtr_ht}.}
    \label{fig_born_ht}
\end{figure}

In Fig.~\ref{fig_subtr_ht}, we illustrate the effect of this prescription (solid black line): at large temperatures, much larger than any mass, it coincides with the LPM curve, as the $1\leftrightarrow2$ rate is only caused by
the soft interactions with the plasma. In the non-relativistic regime,
on the other side of the figure, the LPM curve in dotted green approaches
the collinear Born curve in dashed blue. 
As can be gleaned from Eq.~\eqref{collborn}, for small values of $T/M$,
$k_0$ approaches $M$ and $\mathrm{Im}\Pi_R^\mathrm{LPM\,Born}\propto (M_\eta^2-M^2) T/M$.
This behaviour, reflected in the growth of the blue curve at
small $T$, is unphysical: the proper Born evaluation, without the
collinear approximation, shows that in this region $p_\mathrm{min}
\approx p_\mathrm{max}$,  as $k_0\gg k$; this can be read off
Eq.~\eqref{boundaries_thermal_born}, and causes the full Born rate to run out
of phase space and approach zero.
Upon performing the
procedure in Eq.~\eqref{sub_pres}, the unphysical growth
at low $T$ is removed and our best-effort curve in solid black
is thus very close to the
full Born of Eq.~\eqref{fullborneta} in dash-dotted red. 
In the ultrarelativistic regime
both Born curves vanish, as the vacuum+thermal $\mathcal{M}_\eta$ is smaller
than $m_q+M$ for this set of parameters.

Finally, in order to see the effect induced by neglecting the thermal quark mass in the 
full Born rate, e.g. in using Eq.~\eqref{vacuum_born}, we plot it besides the other
curves, for the same set of parameters, in Fig.~\ref{fig_born_ht}. It
shows how neglecting the quark thermal mass would have a significant impact in the peak
region for $T\sim M$ and for larger temperatures as well.

\subsection{$2 \to 2$ scatterings}
\label{2to2_rate}
In this section we deal with the $2 \to 2$ scattering processes that contribute to the production of DM fermions. For this model, they have been considered  in ref.\cite{Garny:2018ali}, though for $T\lesssim M$
and possibly with some limitations due to the lacking of a proper handling of the IR divergences in some of the diagrams.  
At high temperatures, the $2 \to 2$ scatterings are expected to be of the same order of the LPM rate. When approaching temperatures $T \siml M_\eta$, we shall take into account a phenomenological switch off as described in Sec.~\ref{LT_limit}, in order to use our results for the determination of the dark matter energy density.
\begin{figure}[t!]
    \centering
    \includegraphics[scale=0.5]{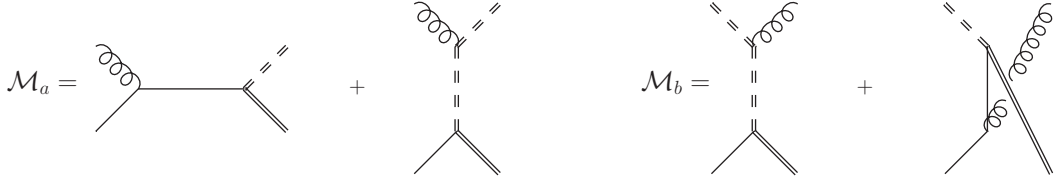}
    \caption{The diagrams contributing to the $2 \to 2$ processes $g \, q \to  \eta \chi$ and $ \eta^\dagger q \to g \chi$ are shown. The same diagrams with a U(1)$_Y$ gauge boson that replaces the QCD gluon (curly line) contribute as well.}
    \label{fig:2to2_ab}
\end{figure}

\begin{figure}[t!]
    \centering
    \includegraphics[scale=0.5]{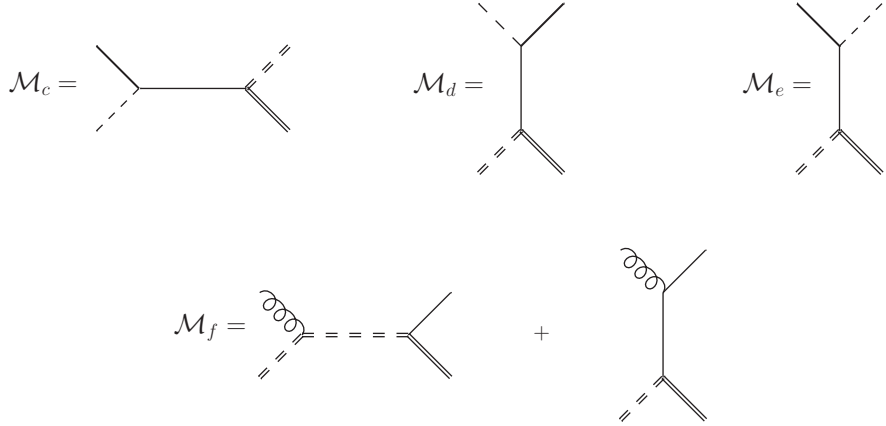}
    \caption{The diagrams contributing to the $2 \to 2$ processes  $Q\phi \to  \eta \chi$,  $\phi \eta^\dagger \to \bar{Q} \chi$, $Q \eta^\dagger \to \phi^\dagger \chi$ and $g  \eta^\dagger \to \bar{q} \chi$. The thick solid line stands for a SU(2)$_L$ left-handed quark doublet, the Higgs doublet is depicted with a dashed line. The same diagrams with a U(1)$_Y$ gauge boson that replaces the QCD gluon are considered as well.}
    \label{fig:2to2_cd}
\end{figure}

The diagrams of the $2\to 2$ processes are collected in Fig.~\ref{fig:2to2_ab} and \ref{fig:2to2_cd}. 
In the case of the $2\to 2$ scatterings,  we exploit 
Eq.~\eqref{diff_rate_eq} and the leading-order equivalence
between $\mathrm{Im}\Pi_R$ and the Boltzmann equation approach 
for $k\sim \pi T$, the hard momentum region. As
we shall see, there are subtleties in this equivalence relating
to soft ($q\sim g T$) momentum exchange. 
For $k \sim \pi T$ and for hard momentum exchange, 
the leading-order contribution to Eq.~(\ref{diff_rate_eq}) from the $2 \to 2$ processes is 
\begin{eqnarray}
  \label{kin_thy}
  \dot{f}_\chi (k)&=& n_\mathrm{F}(k) \Gamma(k) \bigg\vert_{2\leftrightarrow 2}^\mathrm{hard}  \, + \cdots
  \nonumber 
  \\
  &=&\frac{1}{4k}
  \int \! {\rm d}\Omega^{ }_{2\leftrightarrow2} \sum_{abc} 
  \Bigl\vert\mathcal{M}^{ab}_{c\chi}
  (\bm{p}_1,\bm{p}_2;\bm{k}_1,\bm{k})\Bigr\vert^2 
  f^{ }_a(p^{ }_1)\,f^{ }_b(p^{ }_2)\,[1\pm f^{ }_c(k^{ }_1)]  \, + \cdots
  \;,
\end{eqnarray}
where we have again neglected $f_\chi (k)$ for the dark matter appearing in the final state and the ellipses stand for other production processes, such as the $1\leftrightarrow 2$ channels. The $2 \to 2$ phase space measure  is the product of $\int_{\bm{q}} 1/(2 E_q)$ --- for $\bm{q}\in \bm{p}_1,\bm{p}_2,\bm{k}_1$ --- with
$(2\pi)^4 \delta^{(4)}(\mathcal{P}_1+\mathcal{P}_2-\mathcal{K}_1-\mathcal{K})$ and has dimensions of energy squared. The sum runs over all $abc\in$ SM+$\eta$  
particle and antiparticle degrees of freedom  
and thus over all $ab\to c \chi$ processes listed in Figs.~\ref{fig:2to2_ab} and \ref{fig:2to2_cd}, and
$\vert\mathcal{M}^{ab}_{c\chi}\vert^2$ 
is the corresponding matrix element squared,
summed over all degeneracies of each species.
For the SM in the symmetric phase, these are 
spin, polarization, colour, weak isospin and generation. 
For $k\sim \pi T\gg M_\eta,M$ the contribution of thermal 
and in-vacuum masses is suppressed, 
so the external states can be considered massless. 
The prefactor $1/(4k)$ is a combination of $1/(2k)$ from 
the phase space measure and $1/2$ for 
the symmetry factor for identical initial state particles; 
in the cases where $a\ne b$ this factor
is compensated for by their being counted twice
in the sum over $abc$. There is no degeneracy factor for the $\chi$,
as we count the two helicity states as the particle and the antiparticle when considering the number density for $\chi$ and, consistently, we do not include processes producing a $\bar\chi$ in Eq.~\eqref{kin_thy}. 
As can be read off of Eq.~(\ref{defgamma}), we can use the identification $k\Gamma(k) = \vert y\vert^2\textrm{Im}\Pi_R$ ($k^0 \approx k$ at high temperatures), and we write the result of the $2 \to 2$ scatterings as follows 
\begin{eqnarray}
n_\mathrm{F}(k) \textrm{Im}\Pi_R \bigg\vert_{2\leftrightarrow2}^\mathrm{hard} &=&
\frac{1}{2} N_c\bigl( Y_q^2g_1^2+ C_F g_3^2+ \vert h_q\vert^2\bigr)\int \! {\rm d}\Omega^{ }_{2\to2} \bigg\{
   \nonumber\\
     \label{bbf}
   &+&n_\mathrm{B}^{ }(p_1)\,n_\mathrm{B}^{ }(p_2)\,[1-n_\mathrm{F}^{ }(k_1)] 
   \left(\frac{ t}{u}+\frac{ u}{t}\right)\\
   &-&n_\mathrm{F}^{ }(p_1)\,n_\mathrm{B}^{ }(p_2)\,[1+n_\mathrm{B}^{ }(k_1)]
   \left(\frac{ t}{s}+\frac{ s}{t}\right)\label{fbb}\\
   &-&n_\mathrm{B}^{ }(p_1)\,n_\mathrm{F}^{ }(p_2)\,[1+n_\mathrm{B}^{ }(k_1)]
   \left(\frac{ u}{s}+\frac{ s}{u}\right)\label{bfb}\bigg\}
  \;. \hspace*{4mm} 
\end{eqnarray}
We can then exploit the relabeling ${\bm p}_1\leftrightarrow {\bm p}_2$ to
change some $u$ into a $t$,
yielding  
\begin{eqnarray}
n_\mathrm{F}(k) \textrm{Im}\Pi_R \bigg\vert_{2\leftrightarrow2}^\mathrm{hard} &=& 
N_c\bigl( Y_q^2g_1^2+ C_F g_3^2+ \vert h_q\vert^2\bigr)\int \! {\rm d}\Omega^{ }_{2\to2} \bigg\{
   n_\mathrm{B}^{ }(p_1)\,n_\mathrm{B}^{ }(p_2)\,[1-n_\mathrm{F}^{ }(k_1)] 
   \frac{ u}{t}\nonumber \\
   &-&n_\mathrm{F}^{ }(p_1)\,n_\mathrm{B}^{ }(p_2)\,[1+n_\mathrm{B}^{ }(k_1)]
   \left(\frac{ t}{s}+\frac{ s}{t}\right)\label{fbbc}\bigg\}
  \;. \hspace*{4mm} 
  \label{hard_u_u}
\end{eqnarray}
This can be 
further simplified  upon performing the phase-space integrations using the methods in ref.\cite{Ghiglieri:2016xye}.
We find
\begin{eqnarray}
(4\pi)^3 \textrm{Im}\Pi_R \bigg\vert_{2\leftrightarrow2}^\mathrm{hard}&=& 
\frac{2}{k}
N_c\bigl( Y_q^2g_1^2+ C_F g_3^2+ \vert h_q\vert^2\bigr)\bigg\{\int_{k}^\infty dq_+\int_0^k dq_-[n_\mathrm{F}(q_0)+n_\mathrm{B}(q_0-k)]\Phi_{s2}\nonumber\\
&&+\int_{0}^k dq_+\int_{-\infty}^0 dq_-[1-n_\mathrm{F}(q_0)+n_\mathrm{B}(k-q_0)]\Phi_{t2}\bigg\}
\label{hard_phis_phit}
 \end{eqnarray}
 where, in the language of ref.\cite{Ghiglieri:2016xye},  the fermionic $s$-channel and fermionic $t$-channel exchange  functions read 
\begin{eqnarray}
 \Phi_\rmi{$s$2} 
 & = & 
 \biggl\{
  \frac{q}{2} + \frac{T}{q} 
  \bigl[ 
   (k - q_-)(\lnfplus - \lnbminus) + (k - q_+)(\lnfminus - \lnbplus)
  \bigr] 
 \nonumber \\
 & & \; 
   + \, \frac{T^2}{q^2} (2 k - q_0) 
    \bigl(  \libplus{2} + \lifminus{2} - \lifplus{2} - \libminus{2}  \bigr)
 \biggr\}
 \;, \\ 
 \Phi_\rmi{$t$2} 
 & = & 
 \biggl\{
  \frac{T}{q} 
  \bigl[ 
   (k - q_-)(\lnfplus - \lnbminus) + (k - q_+)(\lnfminus - \lnbplus)
  \bigr] 
 \nonumber \\
 & & \; 
   + \, \frac{T^2}{q^2} (2 k - q_0) 
    \bigl(  \libplus{2} + \libminus{2} - \lifplus{2} - \lifminus{2}  \bigr)
 \biggr\}
 \;,\label{phit2}
\end{eqnarray}
where
\begin{eqnarray}
 && 
 \lnfminus \equiv \ln \Bigl( 1 + e^{-|q_-|/T} \Bigr)
 \;, \quad
 \lnfplus \equiv \ln \Bigl( 1 + e^{-q_+/T} \Bigr)
 \;, \label{not1} \\
 && 
 \lnbminus \equiv \ln \Bigl( 1 - e^{-|q_-|/T} \Bigr)
 \;, \quad
 \lnbplus \equiv \ln \Bigl( 1 - e^{-q_+/T} \Bigr)
 \;, \\
 &&
 \lifminus{i} \equiv {\mbox{Li}}^{ }_i \Bigl( - e^{-|q_-|/T} \Bigr)
 \;, \quad  \hspace*{3.2mm}
 \lifplus{i} \equiv {\mbox{Li}}^{ }_i \Bigl( - e^{-q_+/T} \Bigr)
 \;, \\
 && 
 \libminus{i} \equiv {\mbox{Li}}^{ }_i \Bigl( e^{-|q_-|/T} \Bigr)
 \;, \quad \hspace*{6mm}
 \libplus{i} \equiv {\mbox{Li}}^{ }_i \Bigl( e^{-q_+/T} \Bigr)
 \;, \label{notation}
\end{eqnarray}
and where
\begin{equation}
 q^{ }_{\pm} \; \equiv \; \frac{q_0 \pm q}{2} 
 \;. 
 \label{qpm}
\end{equation}
As pointed out in ref.\cite{Besak:2012qm}, whenever the matrix element squared is proportional to $s/t$ or $u/t$, there are leading order
contributions from both hard and soft momentum transfer. For this model too, such structure do appear in Eq.~\eqref{hard_u_u} and consequently in Eq.~\eqref{phit2}. 
Hence, while $\Phi_{s2}$ remains finite in the whole integration region,  the $t$-channel function $\Phi_{t2}$ has a non-integrable divergence at $q \ll k$. It is exactly this ill behaviour that signals the leading-order sensitivity to soft fermion exchange and the need to implement a proper HTL resummation to obtain a finite and physical result. As detailed in ref.\cite{Ghiglieri:2016xye}, we implement the subtraction from the $t$-channel term in Eq.~(\ref{hard_phis_phit}) of the divergent soft contribution, so to cancel the pathological behaviour at $q \ll k$, and add the proper HTL resummed contribution. The implementation of these steps   provide us with 
\begin{eqnarray}
\textrm{Im}\Pi_R^{2\leftrightarrow 2} & = & 
 \frac{2}{(4\pi)^3k}
 \int_{k}^{\infty} \! {\rm d} q_+ \int_0^{k} \! {\rm d} q_- 
 \Bigl\{ 
  \bigl[ n_\mathrm{F}{}(q_0) + n_\mathrm{B}{}(q_0 - k)  \bigr] 
  N_c\bigl( Y_q^2g_1^2+ C_F g_3^2+ \vert h_q\vert^2\bigr) \,  \Phi_\rmi{$s$2} 
 \Bigr\} 
 \nonumber \\
 & + & 
 \frac{2}{(4\pi)^3k}
 \int_{0}^{k} \! {\rm d} q_+ \int_{-\infty}^{0} \! {\rm d} q_-  
 \Bigl\{ 
  \bigl[1 - n_\mathrm{F}{}(q_0) + n_\mathrm{B}{}(k - q_0) \bigr]   
 N_c\bigl( Y_q^2g_1^2+ C_F g_3^2+ \vert h_q\vert^2\bigr)  \,\Phi_\rmi{$t$2} 
 \nonumber\\
 & & 
  -  \, \Bigl[n_\mathrm{B}{}(k) + \frac12\, \Bigr]
 \,  
 N_c\bigl( Y_q^2g_1^2+ C_F g_3^2+ \vert h_q\vert^2\bigr) \,  \frac{k \pi^2 T^2}{q^2}
 \, \Bigr\} 
 \nonumber \\
 & + & 
 N_c\frac{m_q^2}{16\pi} \, \Bigl[n_\mathrm{B}{}(k) + \frac12\, \Bigr] 
 \, \ln\left(1+ \frac{4k^2}{m_q^2} \right)
 \; + \; 
 \mathcal{O}\Bigl( \frac{m_q^4}{k^3} \Bigr)
 \;. \label{direct_full}
\end{eqnarray}
As discussed e.g.~in \cite{Ghiglieri:2016xye,Ghiglieri:2020mhm}, we keep the argument of the logarithm in Eq.~(\ref{direct_full}) in that
form, without resorting to the simplification valid for $k\gg m_q$, as it
is a better (non-negative) prescription at small momenta $k\ll T$, where this calculation
is no longer valid.

\subsection{Low-temperature limit}
\label{LT_limit}
In preparation for the next Sec.~\ref{numerical_results}, where we shall extract the dark matter energy density, we need to complement the treatment of the high-temperature processes in order to follow the entire production process down to smaller temperatures. The main point here is that, while the universe cools down, the dynamics of the production processes is increasingly affected by the in-vacuum masses.  As we assumed $k\sim \pi T \gg M,M_\eta$, the calculations presented in Secs.~\ref{LPM_rate} and \ref{2to2_rate} cease to
be valid for $T\lesssim M,M_\eta$, where the full Born term
of Eq.~\eqref{fullborneta} is the leading-order term and subleading 
corrections, as we mentioned earlier, are unknown in this relativistic regime. In order to switch off such rates progressively when
approaching this region in our phenomenological analyses, we follow the recipe
of ref.\cite{Ghiglieri:2016xye}: we shall multiply $(\mathrm{Im}\Pi_R^{\hbox{\tiny LPM}} -\mathrm{Im}\Pi_R^{\hbox{\tiny Born LPM}})$ and $\mathrm{Im}\Pi_R^{2\leftrightarrow2}$ by a factor $\kappa(M_\eta)$, which is obtained
by taking the $\eta$ boson susceptibility normalised by its massless limit.\footnote{%
In the $2\leftrightarrow2$ case the prescription adopted here coincides with that of
\cite{Ghiglieri:2016xye}. In the $1\leftrightarrow2$ case the different mass pattern
made it so that, in the broken phase, the Born rate (without thermal mass contributions)
was larger than the collinear $1\leftrightarrow2$ rate close to the phase transition,
with the two eventually crossing at lower temperatures
once the collinear $1\leftrightarrow2$ starts its
unphysical increase once the collinear approximation fails, driving the growth of the collinear Born term. In \cite{Ghiglieri:2016xye} one simply transitions from the collinear $1\leftrightarrow2$ rate to the Born rate at their crossing, see Fig.~3 there.} The susceptibility factor reads
\begin{equation}
    \label{defkappa}
    \kappa(M_\eta)=\frac{3}{\pi^2T^3}\int_0^\infty dp\,p^2\,n_\mathrm{B}(E_\eta)
    [1+n_\mathrm{B}(E_\eta)]\,.
\end{equation}
Even though this suggestions is based on a phenomenological argument rather than a rigorous implementation, it does capture the sensitivity to the largest in-vacuum mass scale in the model, namely $M_\eta$. 
\begin{figure}[t!]
    \centering
    \includegraphics{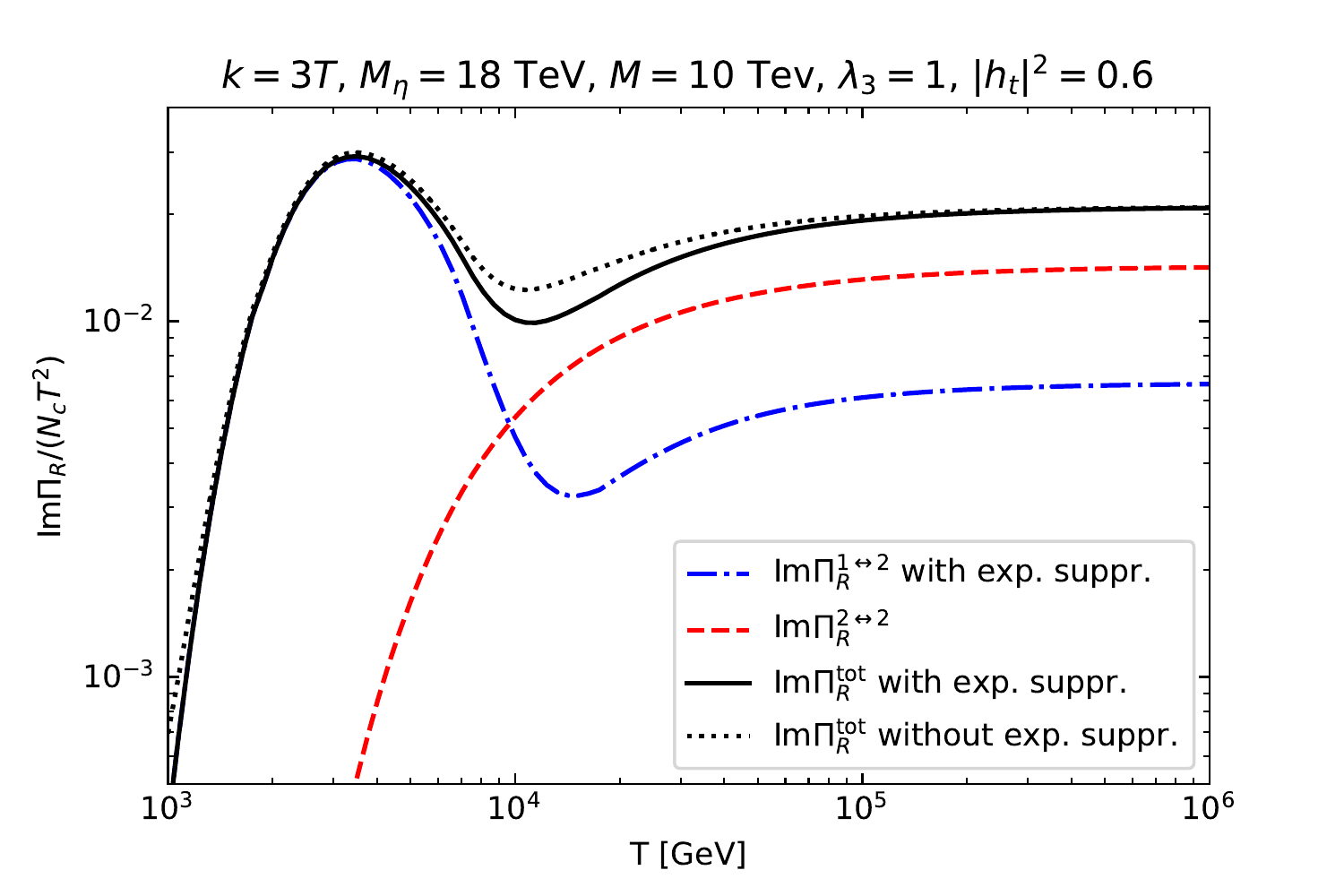}
    \caption{The $1\leftrightarrow2$ rate in Eq.~\eqref{sub_pres_2} and $2\leftrightarrow2$ in Eq.~\eqref{direct_full},
times the susceptibility factor $\kappa(M_\eta)$. The solid black line is their sum as in Eq.~\eqref{defpitot} (the dotted-black curve stands for the $1\leftrightarrow2$ without $\kappa(M_\eta)$).}
    \label{fig_12_22}
\end{figure}

In Fig.~\ref{fig_12_22}, we compare and sum 
our final $1\leftrightarrow2$ results with
our $2\leftrightarrow2$ taking into account
the phenomenological switch-off in Eq.~\eqref{defkappa}. We define the total rate as
\begin{equation}
    \mathrm{Im}\Pi_R^\mathrm{tot}=  \mathrm{Im}\Pi_R^{1\leftrightarrow 2}+
\label{defpitot}
     \mathrm{Im}\Pi_R^{2\leftrightarrow 2},
\end{equation}
where the susceptibility factor affect the collinear LPM rates as follows
    \begin{equation}
    \label{sub_pres_2}
    \mathrm{Im}\Pi_R^{1\leftrightarrow 2}=(\mathrm{Im}\Pi_R^\mathrm{LPM}-
    \mathrm{Im}\Pi_R^\mathrm{LPM\;Born})\kappa(M_\eta)+\mathrm{Im}\Pi_R^\mathrm{Born} \, ,
\end{equation}
whereas $ \mathrm{Im}\Pi_R^{2\leftrightarrow 2}$  is given by Eq.~\eqref{direct_full},
times the susceptibility factor $\kappa(M_\eta)$. For this particular
choice of parameters, the $2\leftrightarrow 2$ contribution is approximately
twice as large as the $1\leftrightarrow2$ in the UR regime. In the relativistic 
regime, $T\sim M_\eta, M$, the $1\leftrightarrow2$ and $2\leftrightarrow2$ rates only reflect a phenomenological recipe, while for $T \siml M_\eta,M$ the $1\leftrightarrow2$ curve is dominated by the leading-order
accurate full Born term of Eq.~\eqref{fullborneta}. In Fig.~\ref{fig_12_22} we show the total rate with and without the susceptibility factor $\kappa(M_\eta)$ in the $1\leftrightarrow2$ component, respectively the solid-black and dotted-black curves. One may appreciate how the inclusion of the susceptibility in the $1\leftrightarrow2$ rate affects the total rate when approaching the relativistic regime, and how it helps in avoiding an overestimation of the production process.

As we mentioned previously, going beyond this phenomenological recipe requires
extending the works of 
\cite{Laine:2013vpa,Laine:2013lka,Laine:2013vma,Ghisoiu:2014ena,Ghisoiu:2014mha,Ghiglieri:2014kma,Jackson:2019mop} on the NLO rate in the relativistic regime, $T\sim M_\eta$, and its matching to the UR regime $T\gg M_\eta$, 
to the case where both the ``FIMP'' state ($\chi$) and an ``equilibrium'' one ($\eta$) have non-negligible masses. In the non-relativistic regime, the general-kinematics, 
vacuum-masses Born rate,
Eq.~\eqref{vacuum_born}, represents the leading-order contribution.\footnote{
\label{foot_splitting}
Thermal masses are negligible if $M_\eta^2-M^2\gg g^2T^2$. If the
splitting introduces a new soft scale, this discussion would need
to be modified.}
NLO corrections come both in the form of $2\leftrightarrow2$
and $1\leftrightarrow 3$ \emph{real processes}, with full accounting of the masses of
external and intermediate $\chi$ and $\eta$ states, and of \emph{virtual processes}, 
that is, the interference of the Born amplitude with thermal one-loop corrections thereto.
As shown extensively in the case where only the external ``FIMP'' state is massive 
\cite{Laine:2013vpa,Laine:2013lka,Laine:2013vma}, real and virtual processes are separately
IR divergent and need to be consistently regulated; only their sum is IR safe and physical. 
This should thus caution against computing just the real processes, e.g.  using automated generators such as \textsc{micrOMEGAs} \cite{Belanger:2018sti}, as the resulting expressions would need
to be treated with an unphysical regulator. 

\section{Numerical results for the dark matter energy density}
\label{numerical_results}
In the model under consideration, one finds two contributions to the energy density of  dark matter \cite{Garny:2018ali,Arcadi:2013aba}: the freeze-in mechanism, that dominates at temperatures $T \simg M_\eta$, and the super-WIMP mechanism\cite{Feng:2003xh,Feng:2003uy}, that instead takes place much later in the thermal history at $T \ll M_\eta$. In the latter case, the freeze-out of the $\eta$ particles occurs similarly to a WIMP (despite the leading interactions being driven by the strong coupling $g_3$), and dark matter is produced in the subsequent $\eta$ decays. These will become efficient much later than the chemical freeze-out because of the very small coupling $y \ll1$. Overall, the observed dark matter energy density is given by 
\begin{equation}
    (\Omega_{\hbox{\tiny DM}}h^2)_{\textrm{obs.}}=  (\Omega_{\hbox{\tiny DM}}h^2)_{\textrm{freeze-in}} +  (\Omega_{\hbox{\tiny DM}}h^2)_{\textrm{super-WIMP}} \, ,
    \label{energy_density_tot}
\end{equation}
where $ (\Omega_{\hbox{\tiny DM}}h^2)_{\textrm{obs.}}=0.1200 \pm 0.0012$\cite{Aghanim:2018eyx}. 
In Sec.~\ref{SW_contribution} and \ref{FI_contribution} we study respectively the super-WIMP and freeze-in contribution. 
\subsection{Super-WIMP contribution}
\label{SW_contribution}
The super-WIMP mechanism is based on the late decays of the heavier particle of the dark sector, here $\eta$, and the relic energy density of the colored scalars determines in turn the one of the dark fermions, the actual dark matter, via the relation 
\begin{equation}
    (\Omega_{\hbox{\tiny DM}}h^2)_{\textrm{super-WIMP}} = \frac{M}{M_\eta}   (\Omega h^2)_{\eta} \, .
    \label{super_WIMP_rel}
\end{equation}
It is clear from Eq.~(\ref{super_WIMP_rel}) that an accurate derivation of the colored scalar abundance $(\Omega h^2)_{\eta}$ is key to the determination of the dark matter energy density. 
The dynamics of the freeze-out and later-stage annihilations of colored scalars, as part of a dark matter model, have been subject to many investigations \cite{Edsjo:1997bg,deSimone:2014pda,Ellis:2014ipa,Liew:2016hqo,Kim:2016kxt,Mitridate:2017izz,Biondini:2018pwp,Biondini:2018ovz,Harz:2018csl}. The main outcome is that QCD gluon exchanges induce the so-called Sommerfeld enhancement along with bound-state formation for the non-relativistic colored pairs, thus substantially reducing the population of the $\eta$'s with respect to the free annihilation cross section. Despite the aforementioned studies' focus on the setting where also the fermion $\chi$ is in equilibrium (larger $y$ couplings), the Sommerfeld enhancement and bound-state effects on the scalar pair annihilation stay the same.

In this work, we shall adapt the treatment of the present model as presented in ref.~\cite{Biondini:2018pwp} by taking the limit of very small $y$'s. On the one hand, the pair annihilations happen at the hard scale $M_\eta$, and are encoded in the matching coefficients of a non-relativistic effective field theory \cite{Bodwin:1994jh}. On the other hand, the non-perturbative dynamics is accounted for by the solution of a Schr\"odinger equation that comprises a thermal potential with both a real and an imaginary part. The former include a Debye-screened potential, whereas the latter accounts for
the frequent soft $2 \to 2$ scatterings with plasma constituents. In this framework, one can describe
the two-particle state entering the hard annihilation and account for the dynamical formation of bound states in a thermal bath, without any assumption on the nature of the annihilating
pair. Indeed, both above- and below-threshold states are handled at the same time \cite{Kim:2016kxt,Kim:2016zyy}, and the progressive melting/formation of bound states at finite temperature share many similarities with in-medium heavy quarkonium \cite{Matsui:1986dk}: see\cite{Mocsy:2013syh,Rothkopf:2019ipj}
for recent reviews and \cite{Laine:2006ns,Beraudo:2007ky,Brambilla:2008cx}
for the complex potential. 

The corresponding thermally averaged annihilation cross section, that refers to colored scalar pair annihilation only, reads\footnote{One can take the matching coefficients in ref.~\cite{Biondini:2018pwp}  and set $y \to 0$. Here, we add the contribution that originates from the U(1)$_Y$, which was not taken into account in ref.~\cite{Biondini:2018pwp}. However, the addition is more formal than practical,  the numerical impact being negligible both for the matching coefficient and the thermally averaged Sommerfeld factor (overall of order of few-per-cent with respect to the QCD effects).}
\begin{equation}
    \langle \sigma_\textrm{eff} v \rangle = \frac{c_3 \bar{S}_3 + c_4 C_\textrm{F} \bar{S}_4}{N_c} \, ,
\end{equation}
where 
$c_3$ and $c_4$, at leading order in $1/M_\eta^2$, are 
\begin{equation}
    c_3= \frac{1}{32 \pi M_\eta^2} \left( 2 g_1^4 |Y_q|^4 + \frac{g_3^4 C_\textrm{F}}{N_c} + \lambda_3^2\right)\, , \quad c_4= \frac{g_3^4(N_c^2-4)}{64 \pi M_\eta^2 N_c} \, ,
\end{equation}
and where $\bar{S}_3$ and $\bar{S}_4$ are the generalized thermally averaged Sommerfeld factors that comprise both bound-state and above-threshold effects. We take $\bar{S}_4=1$,  as it was found to slightly deviate from unity (both for weakly and strongly interacting particles \cite{Biondini:2018pwp,Biondini:2017ufr}). The thermally averaged cross section is plugged into a Boltzmann equation 
\begin{equation}
\frac{dn_\eta}{dt} + 3 Hn_\eta =-\langle \sigma_{{\rm{eff}}} v \rangle (n^2_\eta-n^2_{\eta,{\rm{eq}}}) \, ,
\label{BE_gen}
\end{equation}
where $H$ is the Hubble rate of the expanding Universe and $n_\eta$ denotes the number density of colored scalars.
Upon taking the standard definition of the yield $Y_\eta=n_\eta/s$, where $s$ is the entropy density, and the change of variable from time to $z= M_\eta/T$, one can extract the relic energy density as $(\Omega h^2)_{\eta}=Y_\eta(z_{\textrm{final}})M_\eta/(3.645 \times 10^{-9} \textrm{GeV})$. 

With these ingredients at hand, we can explore the super-WIMP contribution to the dark matter energy density in the model parameter space. If the present model is assumed to be the only source of dark matter, then the freeze-in mechanism must complement the super-WIMP production to the observed value according to Eq.~(\ref{energy_density_tot}). Our aim is to map out the region in the parameter space where the super-WIMP contribution is marginal, whereas the bulk of the dark matter comes from the freeze-in. With respect to the analyses carried out in ref.\cite{Garny:2018ali}, we include bound-state effects in our analysis, that further boost the scalar annihilations in addition to the above-threshold Sommerfeld enhancement.  We take the portal coupling  $\lambda_3 \in [0.0,1.0]$, and consider the plane $(M,\Delta M)$ to visualize the dark matter energy density. 
\begin{figure}[t!]
    \centering
    \includegraphics[scale=0.55]{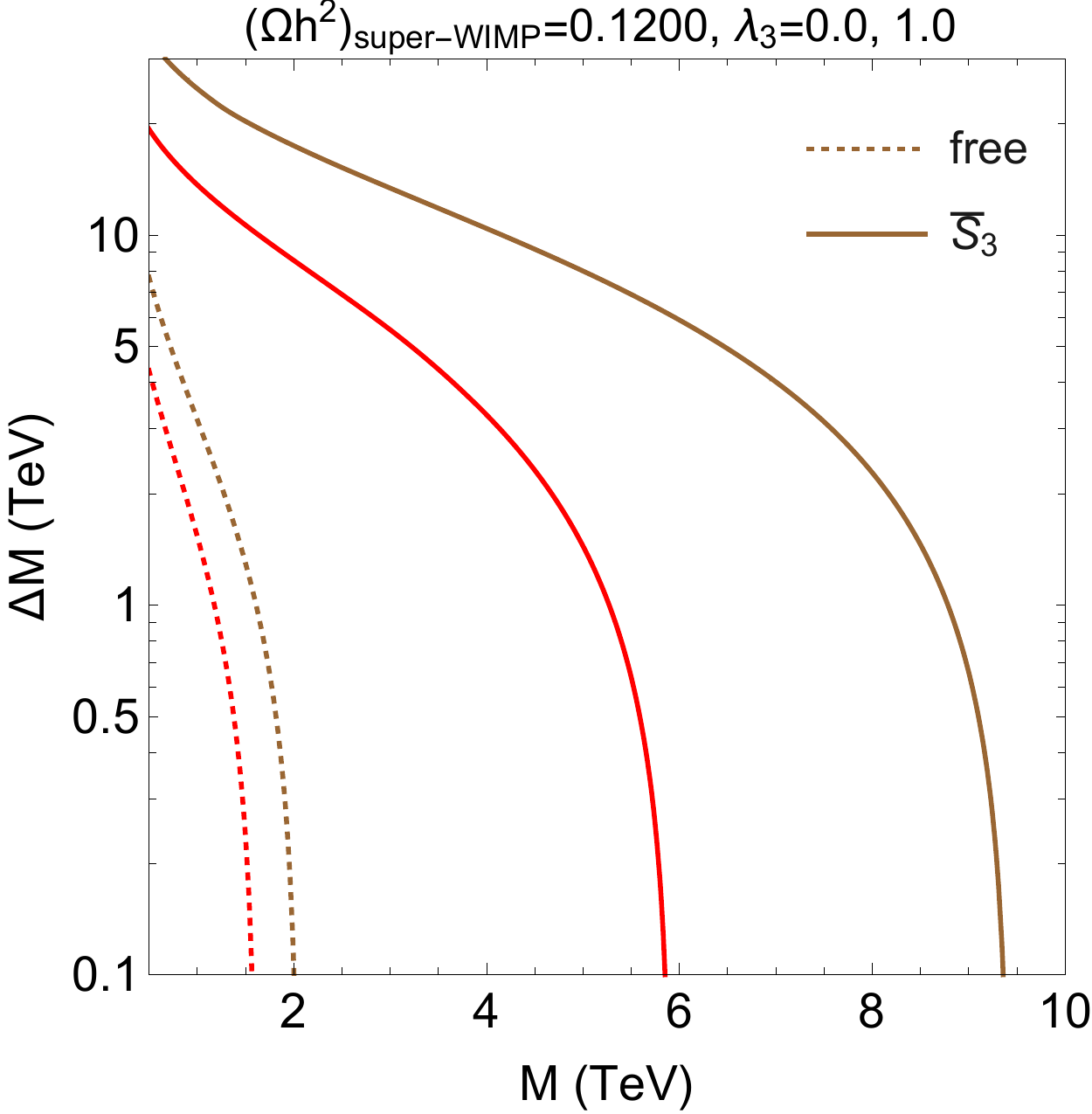}
    \hspace{0.25 cm}
    \includegraphics[scale=0.55]{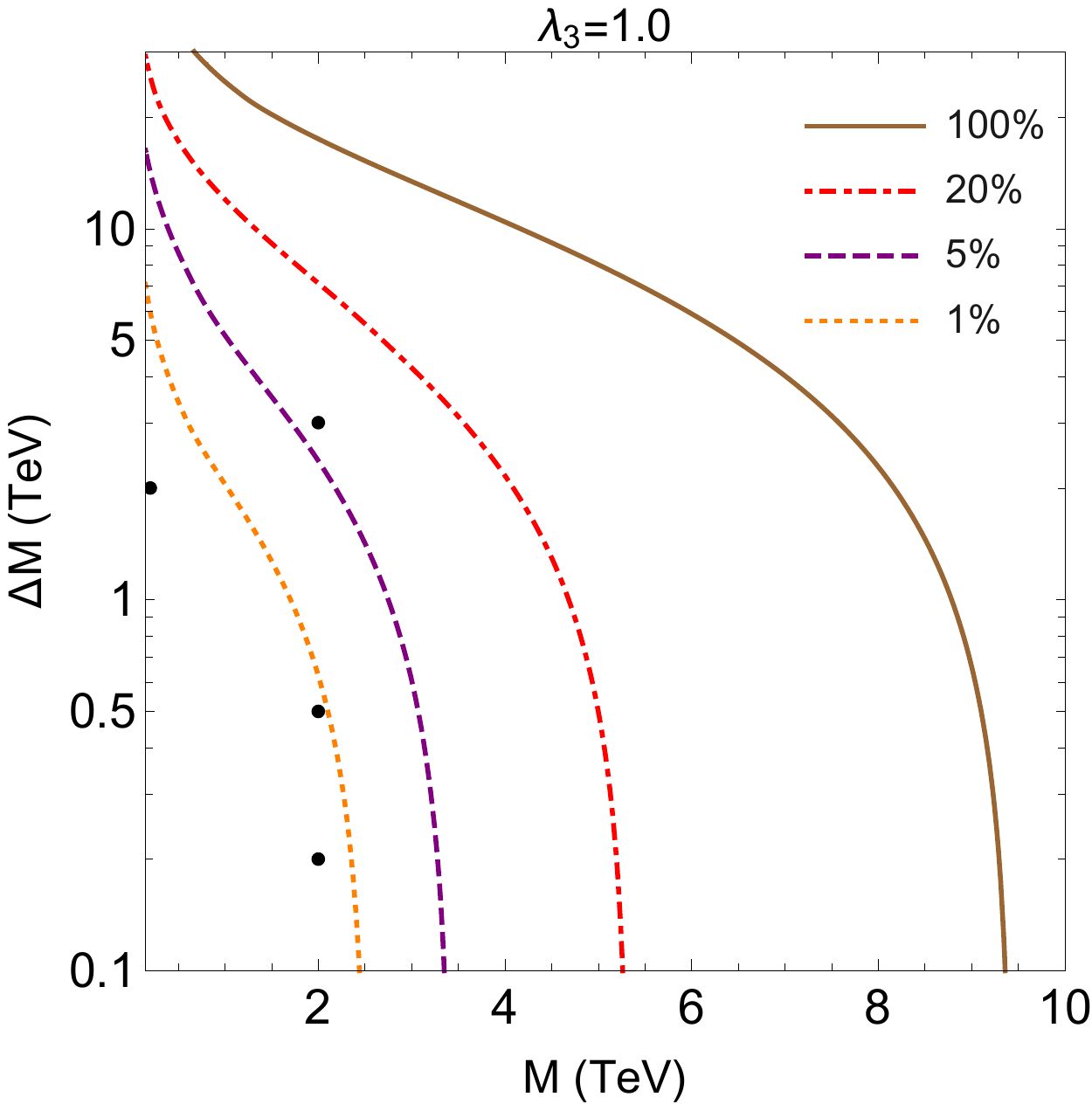}
    \caption{Left panel: the solid and dashed lines reproduce the observed DM relic energy density $(\Omega h^2)_{\textrm{obs.}} = 0.1200$\cite{Aghanim:2018eyx}  in the $(M,\Delta M)$ plane. The solid red ($\lambda_3=0.0$) and brown ($\lambda_3=1.0$) lines are obtained with a thermally averaged cross section comprising Sommerfeld enhancement and bound state effects, whereas the dotted lines are extracted using the free cross section. Right panel: the curves reproduce various fractions of the dark matter energy density upon using the thermally averaged cross section comprising Sommerfeld enhancement and bound state effects, for $\lambda_3=1.0$. We show four benchmark points, black dots, for which we discuss the freeze-in production in Sec.~~\ref{FI_contribution}.}
    \label{fig:super_WIMP}
\end{figure}

The results are given in Fig.~\ref{fig:super_WIMP}, where the dotted curves correspond to the free annihilation cross section ($\bar{S}_i=1$), whereas the solid curves account for the enhanced cross section with Sommerfeld \textit{and} bound-state effects. At a fixed $\Delta M$, the curves shift to larger $M$ values. This is due to a large Sommerfeld factor for the colored scalars so that the same relic density is reproduced for a larger $M_\eta$. In the left panel of Fig.~\ref{fig:super_WIMP}, we set $\lambda_3=0.0$ and $\lambda_3=1.0$. The curves
for $\lambda_3=0.0$ are directly comparable with the results of ref.\cite{Garny:2018ali}, which
did not include that coupling. By comparing to their figure 2, we see that their results, which
only include above-threshold Sommerfeld enhancements, fall between those we obtain from the free cross section and those arising from a full accounting of above- and below-thresholds effects.
Quantitatively, the impact of bound state effects appears sizeable: at $\Delta M=0.1$ TeV 
our largest allowed value for $M$ is more than twice that obtained in \cite{Garny:2018ali}.

In the right panel of Fig.~\ref{fig:super_WIMP}, we instead fix $\lambda_3=1.0$ and indicate different fractions of the observed relic density. For each curve we used the thermally averaged cross section that comprises Sommerfeld enhancement and bound state effects. The black dots correspond to benchmark points in the parameter  space $(M,\Delta M)$ that we shall study in detail in the next section.

A final comment is in order about the extraction of the $Y_\eta(z_{\textrm{final}})$ through Eq.~\eqref{BE_gen}. We integrate the Boltzmann equation up until $z_{\textrm{final}}=10^3$, as a fair compromise between following the evolution of $\eta$ abundance until late times, all the while
staying within a perturbative evaluation of the QCD potential (one expects the transition to a strongly coupled QCD regime around $T \siml 1$ GeV). Another issue is related to a proper form of the rate equation for late times, when the bound-state effects are rather prominent. It appears that neither Eq.~\eqref{BE_gen}, nor the ameliorated version put forward in ref.\cite{Binder:2018znk}, can mitigate the fast depletion of colored scalars for $z \simg 10^3$ \cite{Biondini:2019zdo}. Especially for $M_\eta>1$ TeV, one might stretch the integration to larger $z$ values and still satisfy $T \simg 1$ GeV, so that our results are likely to be conservative. All in all, the super-WIMP contribution for this model could be lower than our findings collected in Fig.~\ref{fig:super_WIMP}.   

\subsection{Freeze-in contribution}
\label{FI_contribution}
In this section we shall use the results of the rates derived in Sec.~\ref{sec_production_spectral} and \ref{production_rate_UR}, and derive the present-day energy density of the fermion $\chi$ from the freeze-in mechanism. We do not perform an entire scan of the parameter space, rather we focus on specific benchmark points (see black dots in Fig.~\ref{fig:super_WIMP} right panel). These belongs to the ``bulk'' region, in the language of \cite{Garny:2018ali}, where freeze-in is expected to dominate DM production.

Taking Eqs.~\eqref{diff_rate_eq} and \eqref{defgamma},
assuming $f_\chi \ll n_{\textrm{F}}(k^0)$ and
recalling that the number density of DM particles is $n_\mathrm{DM}=2\int_{\bf k} f_\chi(k)$,
with the factor of 2 accounting for the two helicity states, and $Y=n_\mathrm{DM}/s$, we have
\begin{equation}
\label{yevolution}
    \frac{d\, Y}{d\,x}=2\frac{\langle \hat\Gamma\rangle}{s},
\end{equation}
where $x\equiv\ln(T_\mathrm{max}/T)$, with $T_\mathrm{max}$ the temperature
where we start the evolution, with $Y(x=0)=0$. 
$\hat O\equiv O/(3c_s^2 H)$, with $c_s^2$
the speed of sound squared and $H$ is the Hubble rate. Finally, $\langle\ldots\rangle\equiv \int_{\bf k} \ldots n_\mathrm{F}(k^0)$.
In what follows, we will use the parametrizations of \cite{Laine:2015kra}
for the speed of sound and entropy and energy densities, the latter entering
the Hubble rate.\footnote{This corresponds to neglecting the extra $2N_c$ bosonic
degrees of freedom contributing to the Hubble rate and entropy density coming
from the $\eta$ scalar at $T\gg M_\eta$, as well as its subsequent freezeout at 
$T\lesssim M_\eta$. The
uncertainty coming from this approximation should be under the 10\% level.}
For what concerns the couplings, we use the one-loop prescriptions of 
Appendix~\ref{sec:running} for the strong coupling and the top-Higgs Yukawa. The SU(2)$_L$
and U(1)$_Y$ couplings are instead kept fixed, as they vary little over the
temperature range we shall consider and are smaller. We choose $\mu=\pi T$
for the renormalisation scale. 
\begin{figure}[t!]
    \centering
    \includegraphics[width=7.4 cm]{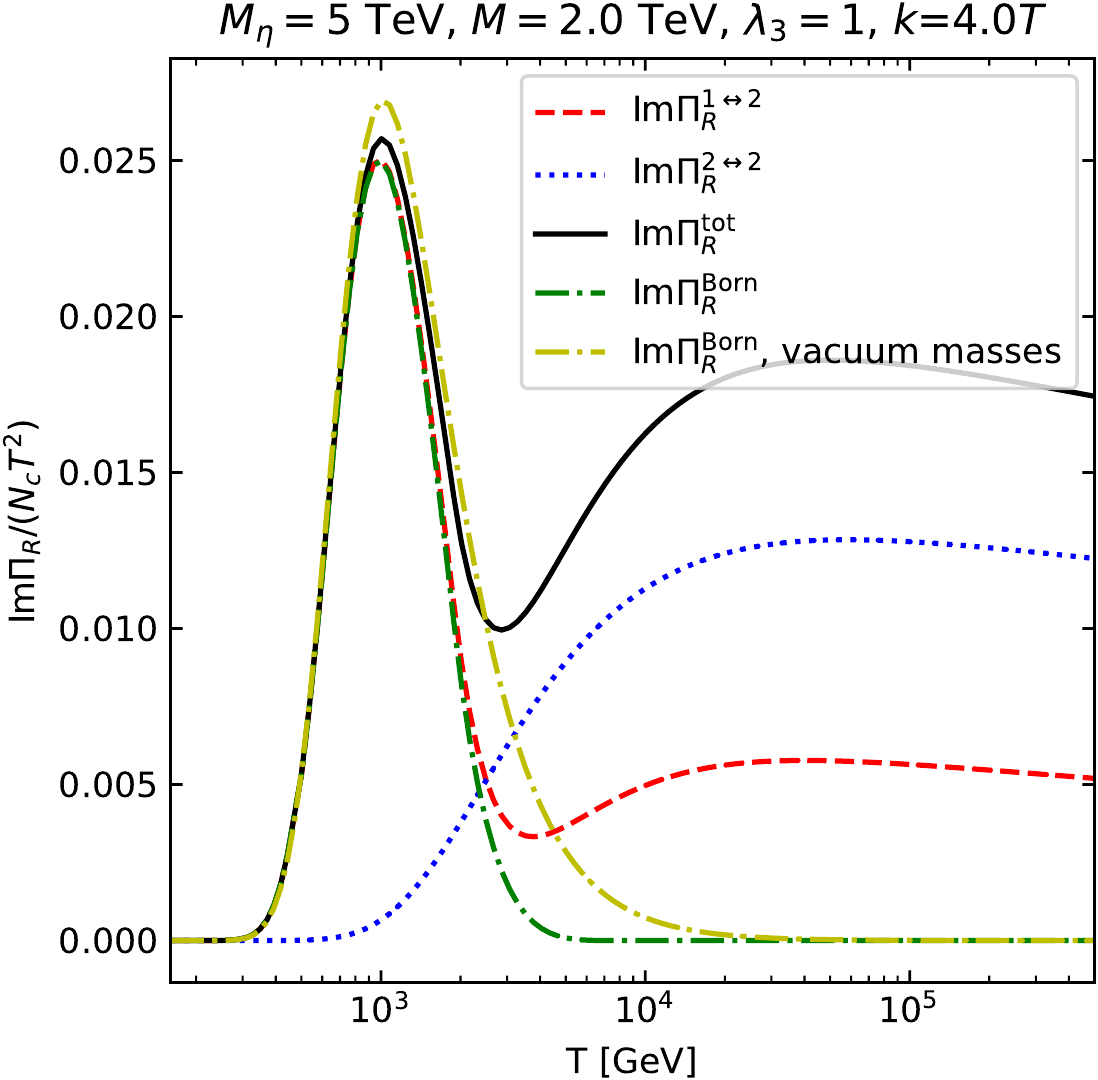}
    \includegraphics[width=7.4 cm]{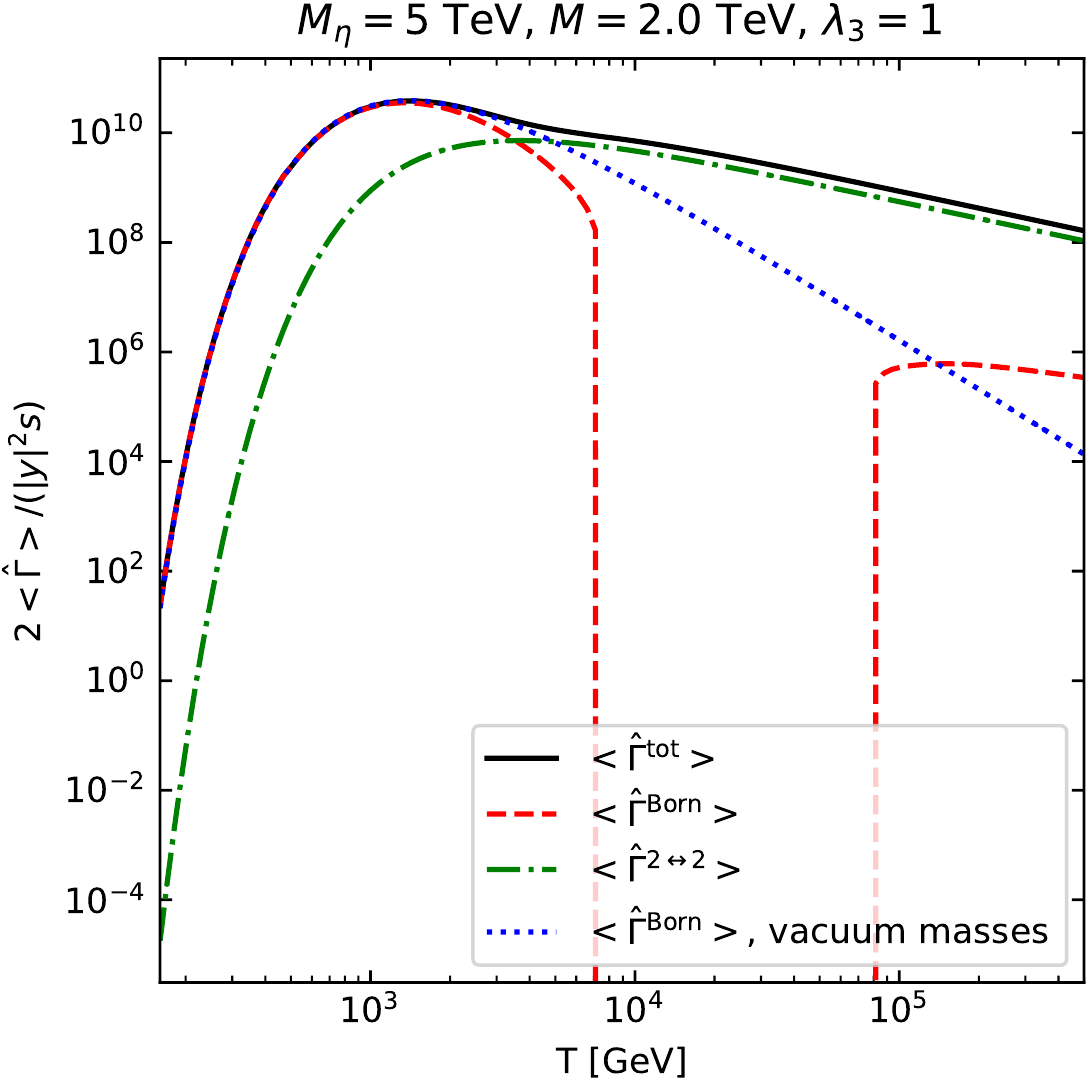}
    \caption{Left: different contribution to the freeze-in rate at  fixed $k/T$. Right: the integrated rate normalized by $3 \vert y\vert^2 c_s^2 H s$. The Born rate in dashed red switches off in the middle of the temperature
    range, because there $M_\eta<m_q+M$, and we do not consider
    in that rate the quark decay contribution of Eq.~\eqref{fullbornq}.}
    \label{fig:IM_Pi_gamma_int_5_2}
\end{figure}

In order to clarify our derivation, we show each step towards the evaluation of the dark matter energy density, starting from the various rates $\textrm{Im}\Pi_R$ to  $(\Omega_{\hbox{\tiny DM}}h^2)_{\textrm{freeze-in}}$. For this purpose, we  choose the benchmark point with $M=2$ TeV, 
$M_\eta=5$ TeV (in-vacuum masses). We consider the up-type quark
coupling to the $\chi$ to be a top, and we fix $\lambda_3=1$.
By the previous analysis, the superWIMP mechanism
is responsible for approximately 7\% of the observed $\Omega_\mathrm{DM}$ (see right plot in Fig.~\ref{fig:super_WIMP}). 

In Fig.~\ref{fig:IM_Pi_gamma_int_5_2} left panel, we show the contributions to $\mathrm{Im}\Pi_R$,
which determine the rate $\Gamma$. In order to appreciate the impact of different approximations, we will track, besides the rate $\Gamma^\mathrm{tot}$
obtained from $\mathrm{Im}\Pi_R^\mathrm{tot}$ in Eq.(\ref{defpitot}), also the rate obtained
from $\mathrm{Im}\Pi_R^\mathrm{Born}$, $\Gamma^\mathrm{Born}$, and the
one where only vacuum masses are used in it, making the top quark massless and
the $\eta$ boson fixed to its vacuum mass (respectively given in Eq.~(\ref{fullborneta}) and (\ref{vacuum_born})). In the right panel of Fig.~\ref{fig:IM_Pi_gamma_int_5_2}, we collect the results of momentum-integrating
the rates, which corresponds (up to the Yukawa $|y|^2$)
to the right-hand side of Eq.~\eqref{yevolution}. It clearly shows
how the total rate, in solid black, behaves as $1/T$ for $T>M_\eta$.
This is consistent with  expectations, since the rate $\Gamma$ there scales approximately 
like $T$, so that $\langle \Gamma\rangle$ scales as $T^4$ and  $\langle \hat \Gamma\rangle/s$ as $m_\mathrm{Pl}/T$, where $m_\mathrm{Pl}= 1.22 \times 10^{19}$ GeV is the Planck mass. Conversely, the Born rates 
behave in this region as $\Gamma\sim (M_\eta^2-M^2)/T$, hence their much steeper power law, as apparent in the case of the massless Born curve (blue-dotted line). 
\begin{figure}[t!]
    \centering
    \includegraphics[width=7.4 cm]{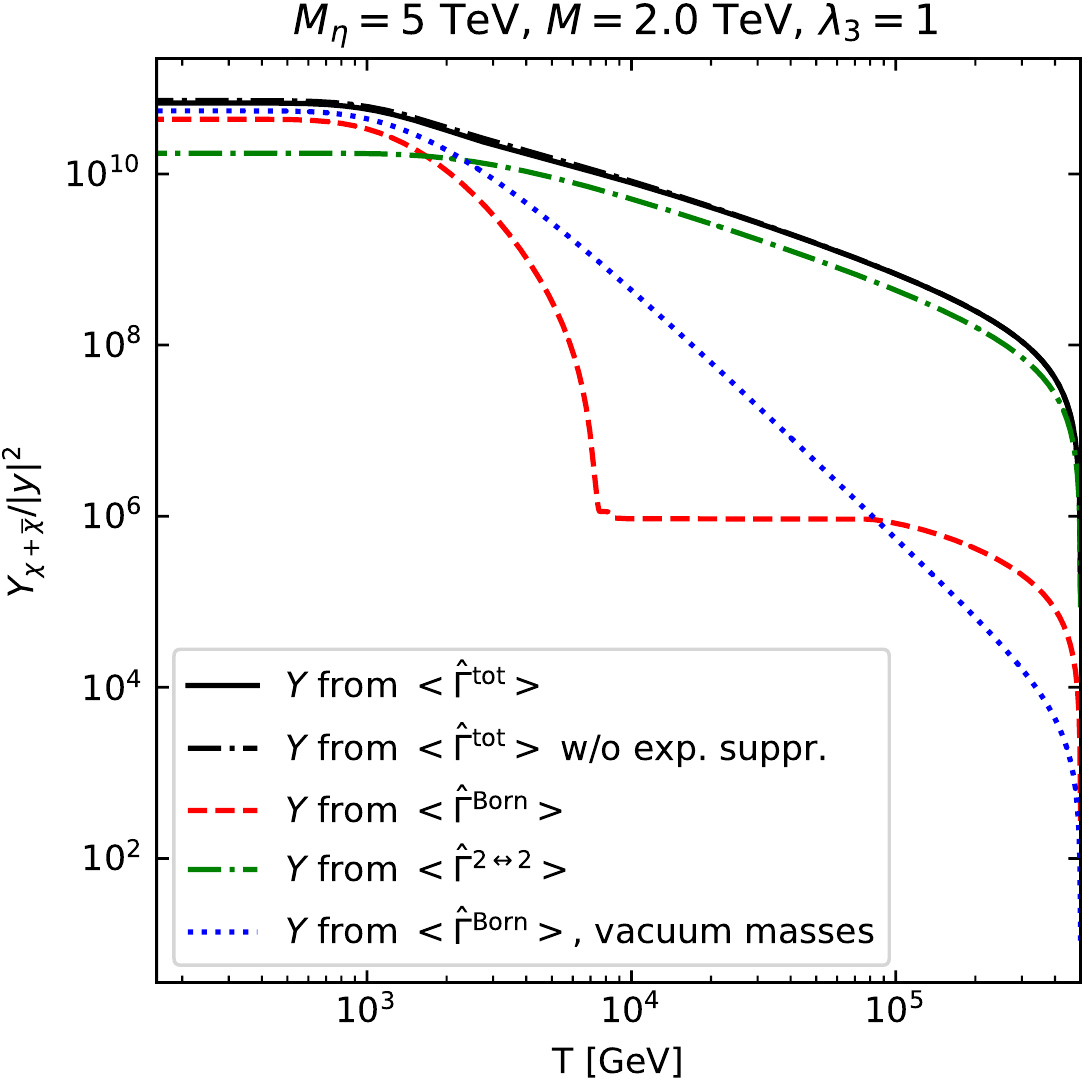}
    \includegraphics[width=7.4 cm]{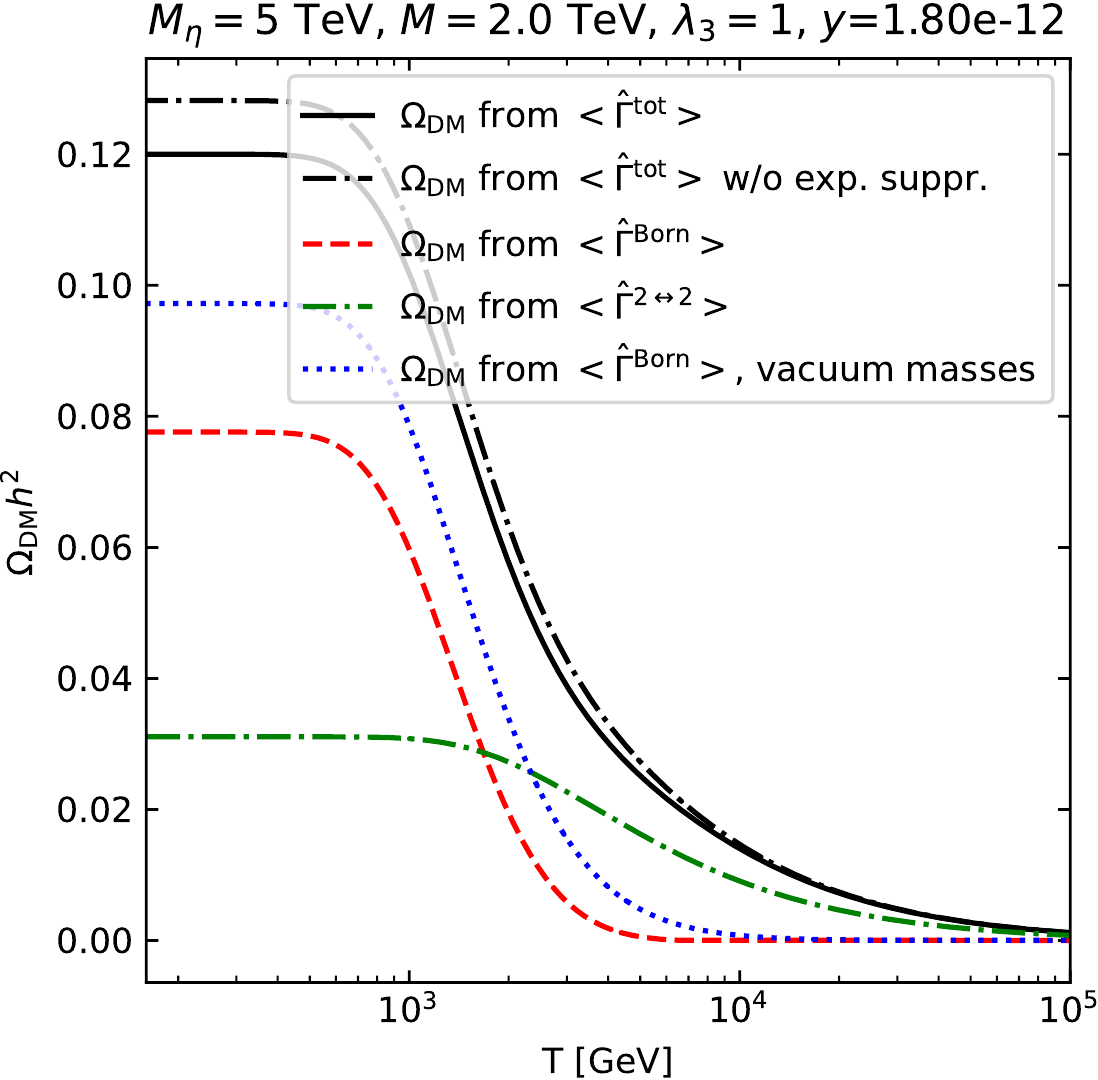}
    \caption{Left: the frozen-in abundance of DM particles divided by the Yukawa coupling squared $|y|^2$. Note that, as the r.h.s. of Eq.~\eqref{yevolution} does not depend on $Y$, the combination $Y/|y|^2$ is most easily determined. The precise value of the initial 
    temperature $T_\mathrm{max}$, here chosen as $T_\mathrm{max}=500$ TeV,
    does not affect appreciably the final abundance, as long as $T_\mathrm{max}\gg M_\eta$
    by a couple of orders of magnitude. Right: the DM energy density
    obtained by tuning the Yukawa coupling to reproduce the observed density (without accounting for the superWIMP contribution) in the case of the $\Gamma^\mathrm{tot}$ rate, black solid line. }
    \label{fig_freezin_abund_5_2}
\end{figure}

Finally, in Fig.~\ref{fig_freezin_abund_5_2} we show the evolution of the yields and DM energy
densities obtained from the freeze-in rates by integrating Eq.~\eqref{yevolution}
in the evolution variable $x=\ln(T_\mathrm{max}/T)$. As the discussion of the previous
paragraph suggests, the approximate $m_\mathrm{Pl}/T$ scaling of 
$\langle \hat \Gamma\rangle/s$ for $T\gg M_\eta$ can be exploited
to estimate how high $T_\mathrm{max}$ needs to be taken to determine
the final abundances to an accuracy that is compatible with the
other uncertainties. In practice, we find that, once $T_\mathrm{max}$
is larger than $M_\eta$ by two to three orders of magnitude, its precise value
does not affect the value we determine for the Yukawa coupling at
the accuracy it is displayed in Fig.~\ref{fig_freezin_abund_5_2}
and the following ones.\footnote{%
Should the universe reheat at a temperature that is lower than
our chosen $T_\mathrm{max}$, the resulting DM abundance would be reduced accordingly.
}
In the right panel, one may see that the correct energy density (neglecting the
superWIMP contribution) is obtained from a coupling which is in the ballpark
of those obtained in \cite{Garny:2018ali}. It also shows how
the Born approximation with vacuum masses corresponds to a $20\%$  reduction of the DM energy density with respect to the full rate (with the Yukawa $y$ tuned to give $(\Omega h^2)_{\textrm{obs.}} = 0.1200$), 
whereas in this case an even larger, $>30\%$ discrepancy, arises from
including thermal masses in the Born rates but neglecting $2\leftrightarrow2$
and medium-induced $1\leftrightarrow2$ processes. The effect of these
is visible already at higher temperatures: at $T=10$ TeV, where they dominate
over the Born processes, they have already produced a \textit{non-negligible}
fraction of the DM abundance, as can be seen from the solid black curve.
It would thus be interesting to have a full NLO analysis in the relativistic regime, along the lines of \cite{Laine:2013lka,Laine:2013vpa,Ghisoiu:2014ena}; we
expect such missing ingredients to also contribute at the tens-of-percent level for this benchmark point. In the case
of smaller $\Delta M/M$, as we shall consider in the following 
subsection, we expect their impact to be more sizeable.
Moreover, we also give the result from dropping the exponential suppression in the $\textrm{Im} \Pi^{1 \leftrightarrow 2}_R$ rate, in order to provide some indication on the theoretical uncertainty one induces by extrapolating the LPM result to the regime $T \siml M_\eta$. In this case we find it to be about a 10\% effect.  

It is also worth remarking how, for $M_\eta=5$ TeV, freeze-in DM production stops well above the electroweak crossover. For lower masses it might become necessary to follow the production rates into the broken phase,
where the methods used here
to determine the thermal contribution to the masses and the $1\leftrightarrow 2$
rates would have to be amended to include the effect of EW symmetry breaking,
as performed in \cite{Ghiglieri:2016xye}. It remains to be understood whether 
any $M_\eta$ exists such 
that is sufficiently low to require that and that is not ruled out
by LHC searches. In this respect, we mention that current searches for detector-stable top-squarks with
the CMS detector exclude masses up to $1250$ GeV \cite{CMS-PAS-EXO-16-036}. Such bound is applicable to the simplified model at hand, whenever the colored scalar decay length $c \tau$ is larger than the detector size (say $\sim 10$ m), where $c$ is the speed of light and $\tau$ is the lifetime of the scalar particle.
\begin{figure}[t!]
    \centering
    \includegraphics[width=7.4 cm]{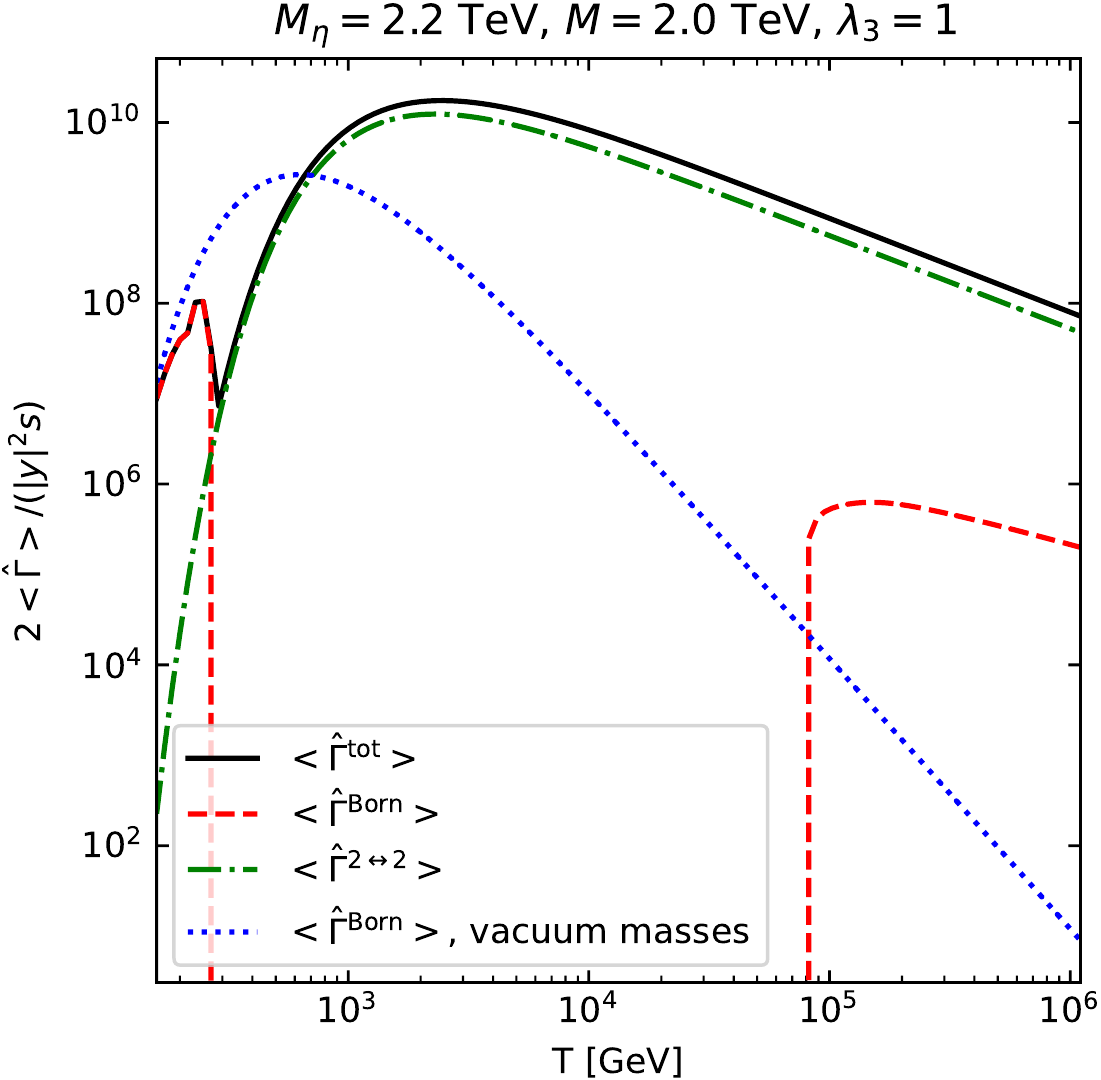}
    \includegraphics[width=7.4 cm]{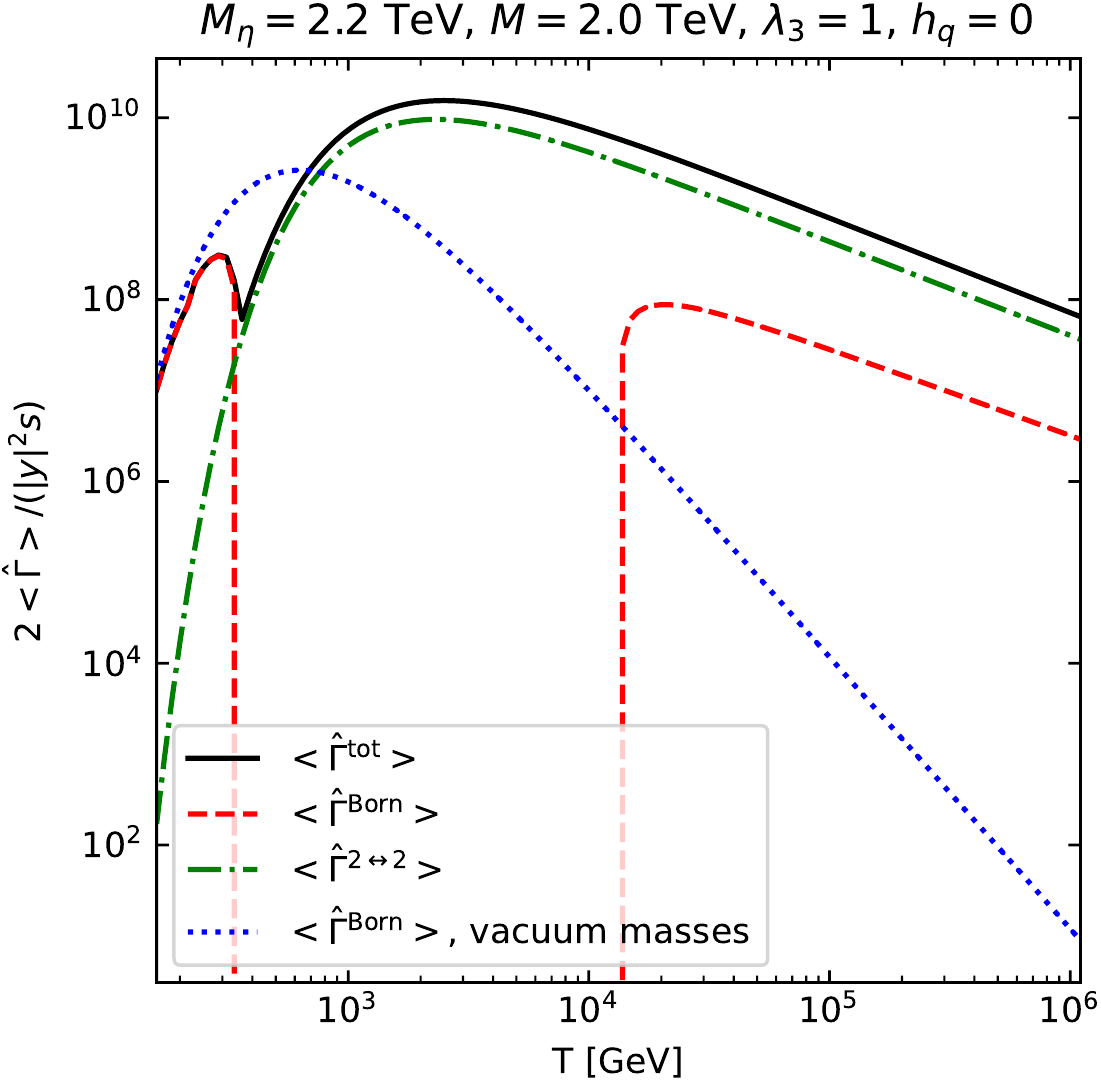}
    \caption{Integrated-in-momentum rates for the benchmark point P3 with $\lambda_3=1.0$. In the left panel, the up-quark Yukawa is $h_t$, running and non-vanishing, whereas the right panel corresponds to $h_q=0$.}
    \label{fig_int_rates_P3_comp}
\end{figure}

\subsubsection{Additional benchmark points}
\label{sec:bench_Ps}
As it is clear from the earlier section, our determination for the production rate in the high-temperature regime results in a sizeable impact on the final dark matter energy density. The relative importance of the total rate (\ref{defpitot}) with respect to the Born rates (\ref{vacuum_born}) and (\ref{fullborneta}) depends on different aspects though. In particular, the in-vacuum mass splitting is rather relevant since it fixes the available room for thermal masses, LPM  and $2 \leftrightarrow 2$ contributions to play a role. As a general trend, the smaller the relative mass splitting $\Delta M /M$, the larger the impact of thermal masses, LPM and $2 \leftrightarrow 2$ processes. This can be understood in terms of the decreasing phase space in the $\eta \to \chi q$ decay process as $\Delta M / M$ gets smaller. However, two couplings can affect this picture, the portal coupling $\lambda_3$ and the up-type quark Yukawa  $h_q$, which give the relative difference between the asymptotic masses of the scalar and the quark (the gauge U(1)$_Y$ and SU(3) contributions are the same, see Eq.~\eqref{asym_the_mass}).

In what follows we consider the three benchmark points as shown in Fig.~\ref{fig:super_WIMP} below the dotted yellow curve. They sit in the region where  freeze-in accounts for almost all the dark matter energy density, namely $(\Omega_\textrm{DM} h^2)_{\textrm{super-WIMP}}/(\Omega_\textrm{DM} h^2)_{\textrm{obs}} <0.01$. We adopt the following convention for the points' labelling: P1 ($M=0.2$ TeV, $M_\eta=2.2$), P2 ($M=2.0$ TeV, $M_\eta=2.5$) and P3 ($M=2.0$ TeV, $M_\eta=2.2$). 
\begin{figure}[t!]
    \centering
    \includegraphics[width=7.4 cm,height=7.4 cm]{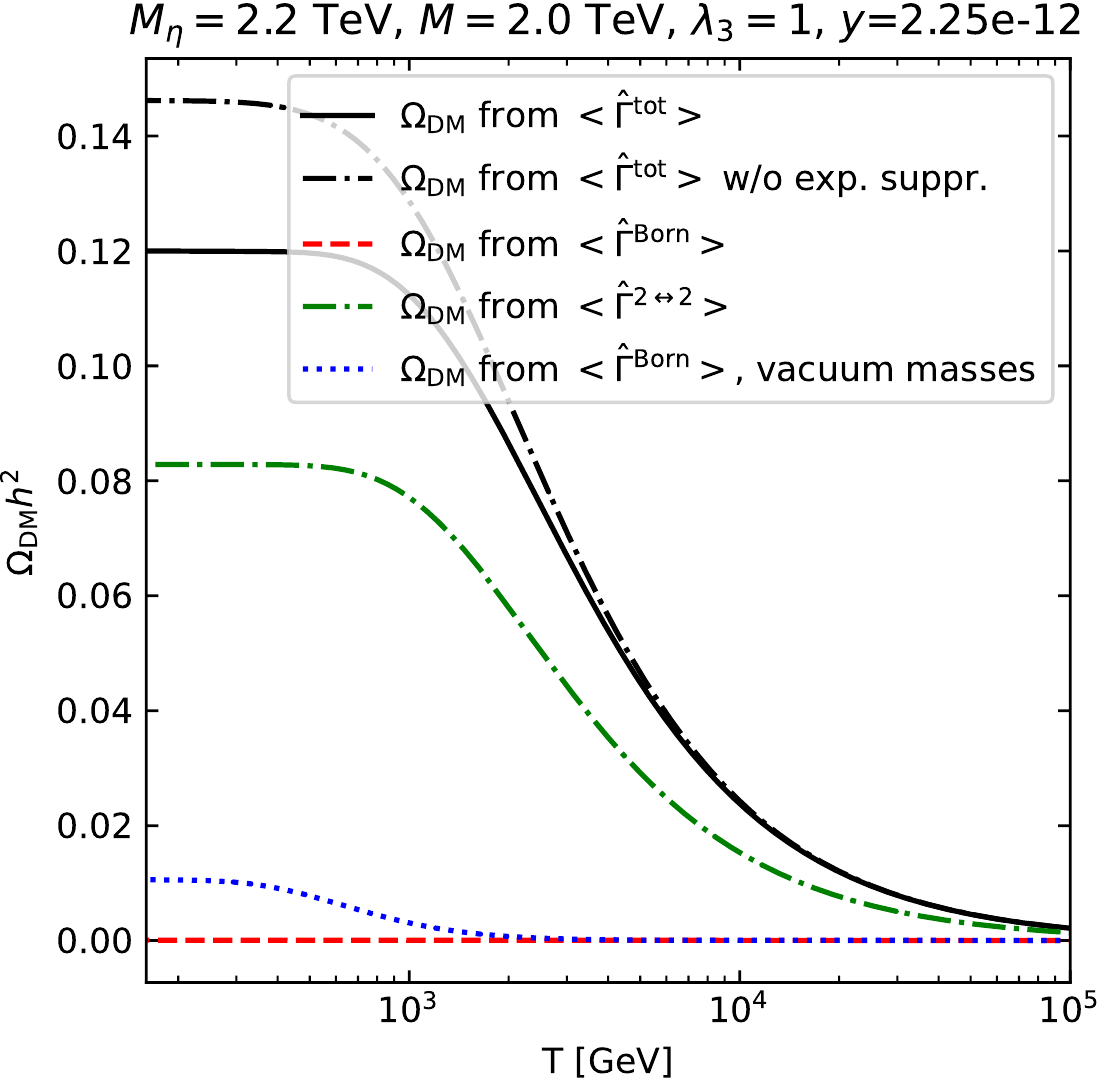}
    \includegraphics[width=7.4 cm,height=7.4 cm]{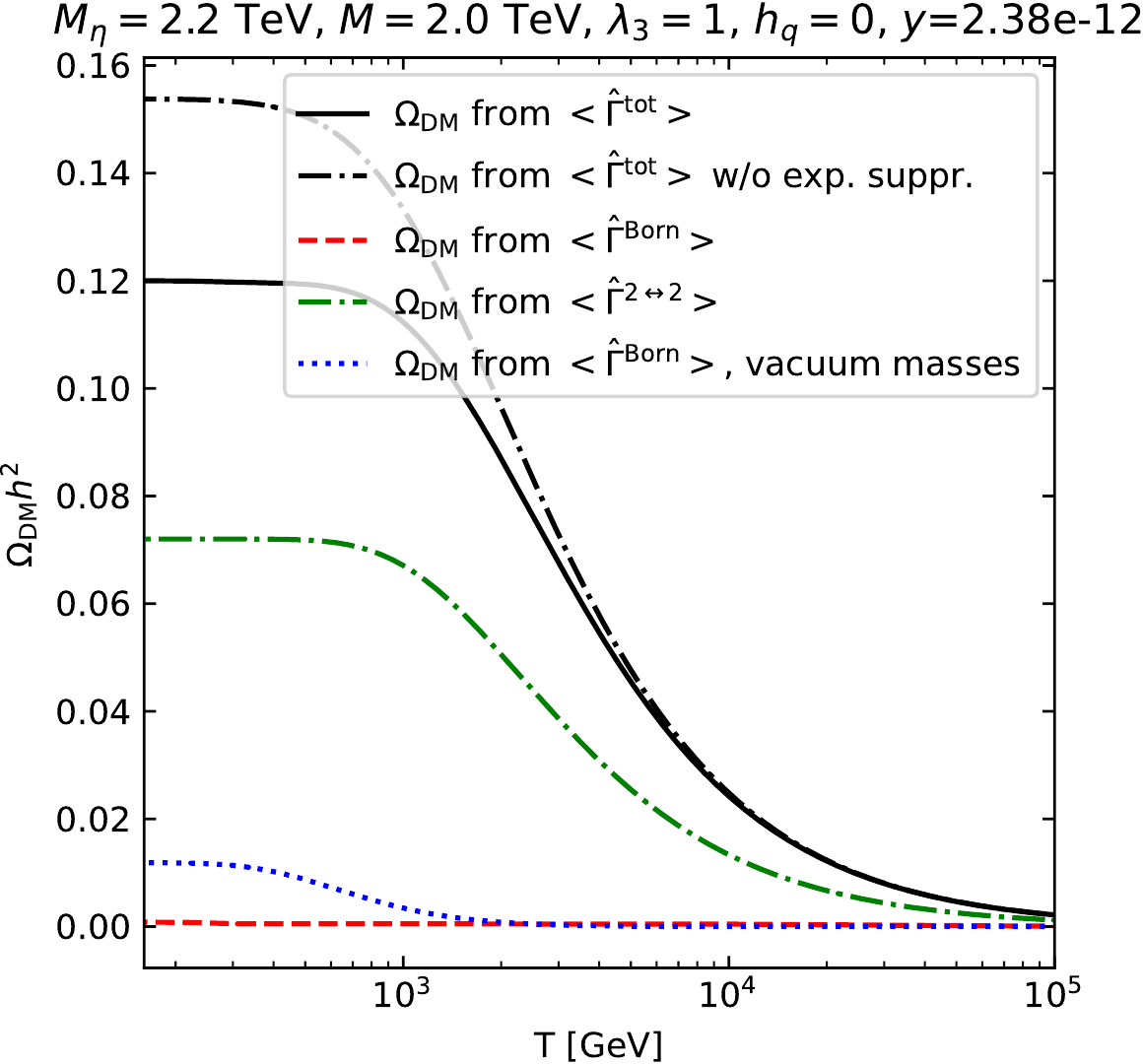}
    \caption{Dark matter energy density for the benchmark point P3 with $\lambda_3=1.0$. In the left panel, the up-quark Yukawa is $h_t$, running and non-vanishing, whereas the right panel corresponds to $h_q=0$.}
    \label{fig_energy_density_P3_comp}
\end{figure}

In order to show some of the effects when changing the coupling combination $(\lambda_3,h_q)$, we consider the benchmark point P3. We display in Fig.~\ref{fig_int_rates_P3_comp} and in Fig.~\ref{fig_energy_density_P3_comp} respectively the momentum-integrated rates and the energy density for $(\lambda_3=1, h_q = h_t\ne 0)$ and $(\lambda_3=1, h_q=0)$. One can appreciate that the $2 \leftrightarrow 2$ contribution diminishes for $h_q=0$, see Eq.~\eqref{direct_full} for the expression of the $2 \leftrightarrow 2$ rate. Then, when the quark thermal mass is due to the gauge contribution only, we see in Fig.~(\ref{fig_int_rates_P3_comp}) that a smaller window within which the Born rate in red-dashed curve vanishes with respect to the case where the Yukawa contribution is instead present. Also the Born rate is larger both in the high-temperature region and around the peak region when $h_q=0$.  Additional comments regarding the energy density curves in Fig.~\ref{fig_energy_density_P3_comp} are in order. First, the dark matter energy density as obtained from the vacuum Born rate accounts only for 10\% of the observed value. When considering the thermal masses in the Born rate, and neglecting the production induced by LPM and $2 \leftrightarrow2$ processes, the corresponding energy density drastically drops to vanishing values, of the order of half a percentage point of the density
produced when accounting for all mechanisms.
Second, the dot-dashed black curve stands for the total rate with unsuppressed LPM contribution. The discrepancy with the solid black line
is larger in the right plot of Fig.~\ref{fig_energy_density_P3_comp} because the $2 \leftrightarrow 2$ contribution is smaller due to $h_q=0$.

\begin{figure}[t!]
    \centering
    \includegraphics[width=7.4 cm,height=7.4 cm]{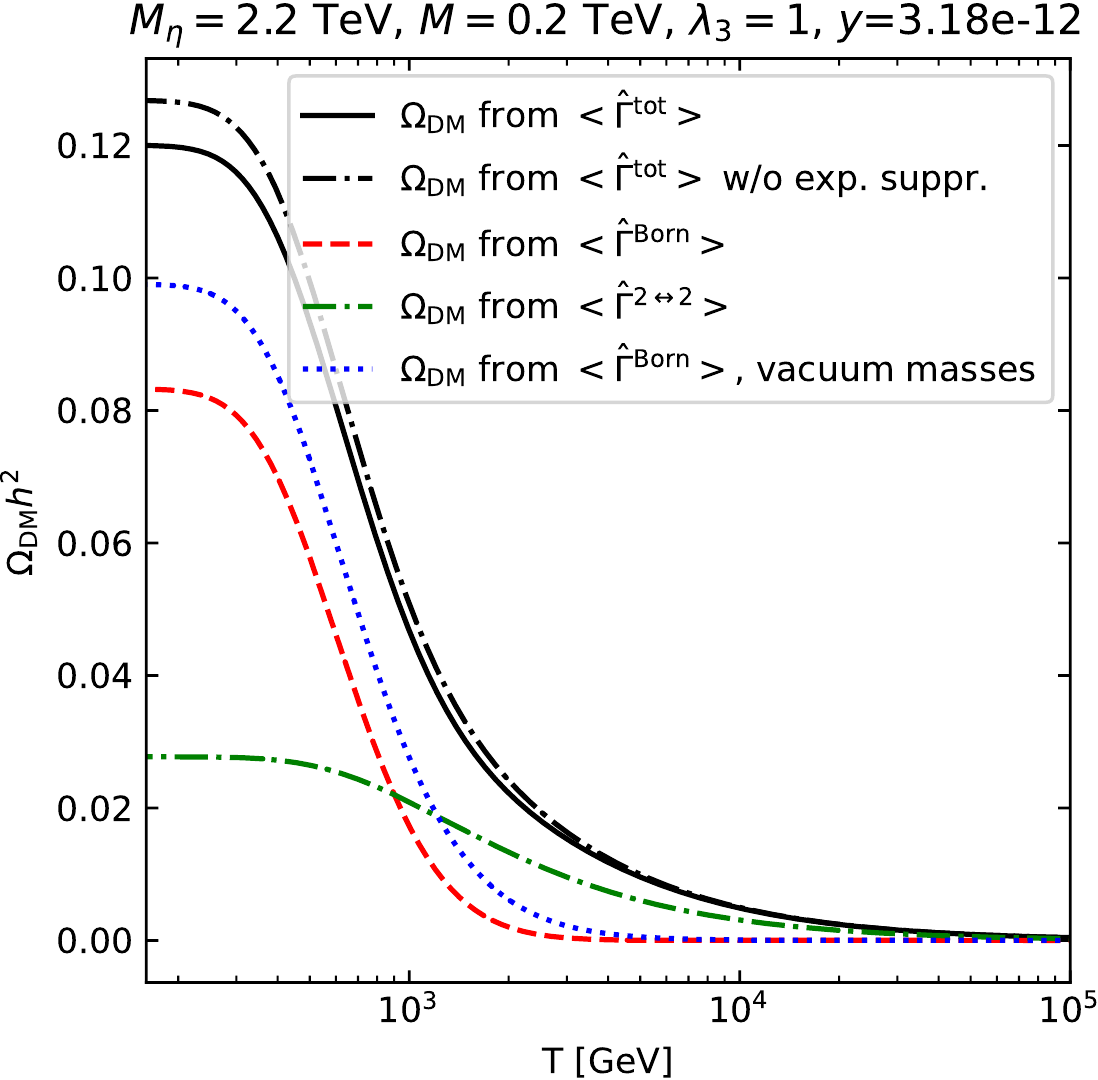}
    \includegraphics[width=7.4 cm,height=7.4 cm]{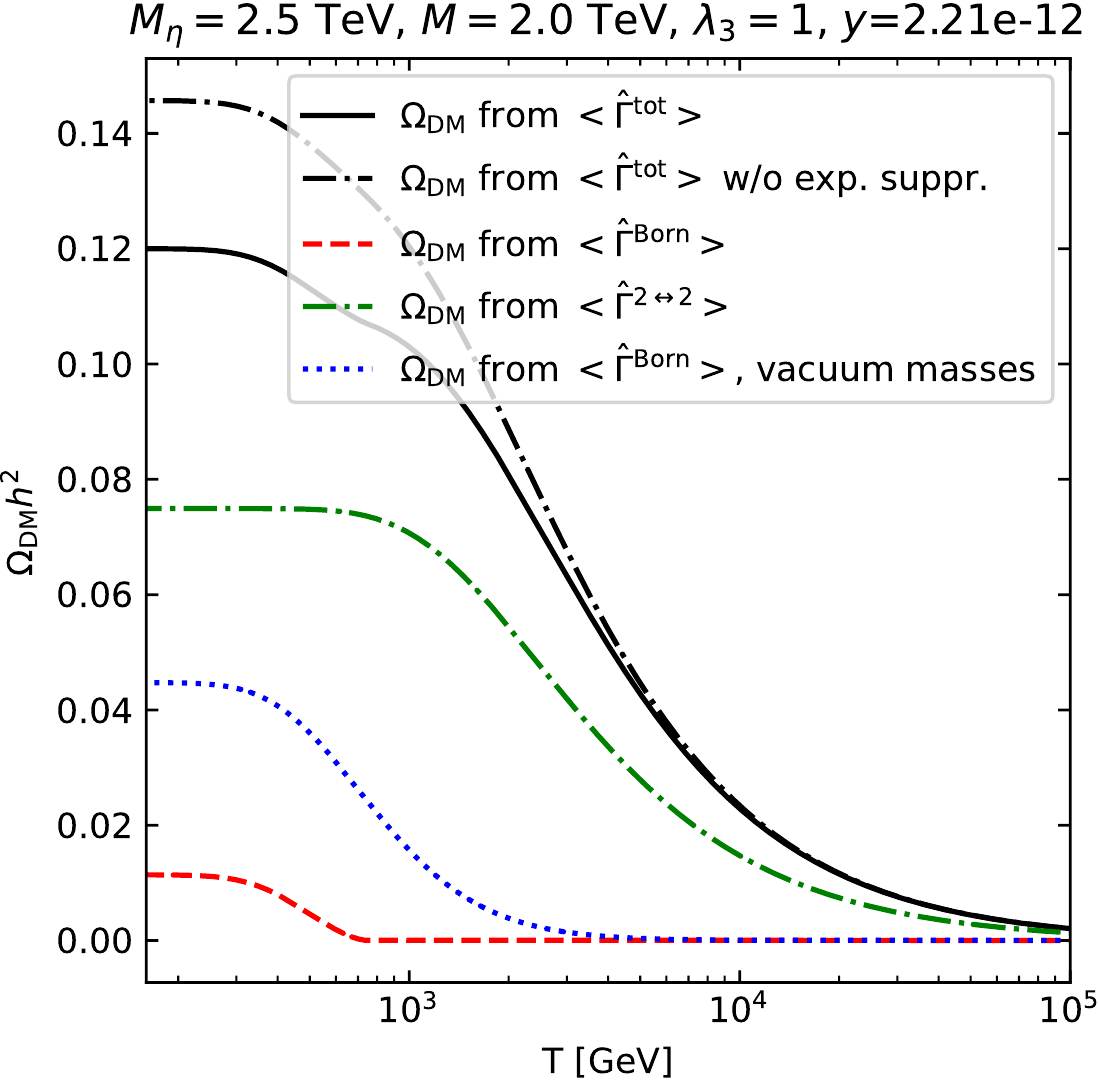}
    \caption{Dark-matter energy density for the benchmark points P1 and P2 with $\lambda_3=1.0$ and running and non-vanishing $h_q=h_t$.}
    \label{fig_energy_density_P2_P1_comp}
\end{figure}

Finally, we show the result for the energy density evolution for the other two benchmark points P1 and P2 in Fig.~\ref{fig_energy_density_P2_P1_comp}. As for the relative mass splitting, one obtains $\Delta M/M =10$ and $\Delta M/M =0.25$ for P1 and P2 respectively, whereas $\Delta M/M =0.1$ for P3. As already mentioned in the general discussion, the effect of the high-temperature dynamics is less important as the relative splitting increases. Nevertheless, for the largest mass splitting considered here, high-temperature contributions still gives a 20\% (40\%) correction to the in-vacuum (thermal-mass included) Born production. 
It is worth noticing that, for the benchmark point P1, the production of the dark matter stops fairly close to the electroweak crossover temperature (see solid-black line in Fig.~\ref{fig_energy_density_P2_P1_comp} left). This calls perhaps for considering the techniques developed in ref.~\cite{Ghiglieri:2016xye} if one allows for smaller $M_\eta$ --- though experimentally $M_\eta > 1250$  GeV \cite{CMS-PAS-EXO-16-036} --- and keeps the relative splitting equally large.

Now that we have extracted the model parameters for the benchmark points in order to reproduce the dark matter energy density, we briefly comment on the constraints from Big Bang Nucleosynthesis (BBN). The tiny $y$ couplings  determine a quite late decay of the scalar mediator population and the corresponding activation of the super-WIMP production. Hence, the energetic decay products are released in the plasma late in the thermal history and can spoil the predictions of the light-element abundance \cite{Jedamzik:2006xz,Kawasaki:2017bqm}. We find the lifetime of the scalar particle to be $\tau_{\textrm{P1}}=0.002$ s,  $\tau_{\textrm{P2}}=0.02$ s, $\tau_{\textrm{P3}}=0.2$ s respectively. For each of the benchmark point we considered the coupling of the colored scalar with a top-like quark, that makes for the largest decay times. Based on the analyses in ref.~\cite{Jedamzik:2006xz}, we checked that the benchmark points do not conflict with the BBN constraints, which exclude energy densities for the colored scalar well above the ones necessary to complement the freeze-in produced fraction with the super-WIMP mechanism.

\section{Conclusions and outlook}
\label{concl_and_outlook}
In this work we have studied the impact of the ultra-relativistic regime on the production of a feebly interacting dark matter particle. As its population accumulates over the thermal history, we inspected thoroughly the temperature window $T \gg M$, which has been previously neglected in the context of dark matter models with renormalizable operators. Here the bulk of the dark matter production is expected to happen for $T \sim M$, so to define what is usually dubbed in the literature as infra-red freeze-in. However, our findings show that this is not always the case: the dark matter population, as produced in the high-temperature regime, can largely contribute, up to $\mathcal{O}(1)$, to the total observed energy density. This is due to very efficient processes that take place in a high-temperature plasma, where all particles can be regarded as effectively massless and a collinear kinematic is established. In particular we addressed multiple soft scatterings, that enhance the $1 \to 2$ decay process and make others viable, as well as $2 \to 2$ scatterings. 

In order to implement such a program, we picked a model with a Majorana fermion dark matter accompanied by a heavier state of the dark sector. The latter shares interactions with the Standard Model and then acts as a mediator between the visible and hidden sectors. Here, the mediator is a scalar field charged under SU(3) and U(1)$_Y$ gauge groups of the Standard Model, and it enters a Yukawa interaction with the dark matter and a right-handed quark. Also, a portal coupling between the scalar mediator and the Higgs doublet appears, whose impact we have addressed in our study. 

After the calculation of thermal (asymptotic) masses and the definition of the production rate in a quantum-field-theoretical fashion at finite temperature, we scrutinized the impact of various effects by taking the Born rate for $\eta \to \chi q$ process with in-vacuum masses as a reference, see Eq.~\eqref{vacuum_born}. First, we include the thermal masses of the scalar and the SM quark in the Born rate, see Eq.~\eqref{fullborneta}. Depending on the choice of the portal coupling $\lambda_3$ and the quark flavor, the phase space for $\eta \to \chi q$ remains closed for a quite extended temperature window with respect to the Born rate with in-vacuum masses. In general, one systematically overestimates the dark matter energy when using the in-vacuum Born rate with respect to the Born rate with thermal masses. Then we derived, and numerically determined, the production rate due to multiple soft scatterings, as mediated by the gauge bosons, that are resummed in the LPM rate, see Eq.~\eqref{LPM_rate}. Here, many reactions can occur: effective and enhanced $1 \to 2$ process $\eta \to \chi q$, effective $1 \to 2$ process $q \to \eta \chi $ and effective $2 \to 1$ process $\eta q \to \chi$. Even though we build up on the recent developments on the production/equilibration rate of Majorana neutrinos in the leptogenesis framework, we include and implement the LPM effect for the first time in the context of freeze-in dark matter. Last but not  least, we included the $2 \to 2$ scattering processes that contribute to the production of dark fermions. Their relevance depends on the choice of the parameters, such as couplings and relative mass splitting. Differently from previous studies on this very model, we treat with care the soft momentum region arising in the $t$-channel quark exchange. We separate the hard and soft region in our derivation, and a finite result is obtained when implementing the HTL resummation for the quark propagator without the need of any ad-hoc regulator, see Eq.~\eqref{direct_full}.  

In order to assess the relative contribution of the super-WIMP and freeze-in mechanism to the present-day dark matter energy density, we included bound-state effects in the determination of the colored scalar yield. The bound-state formation especially boosts the pair annihilations at late times and efficiently depletes the population of colored scalars. With respect to previous estimations \cite{Garny:2018ali}, we find that the super-WIMP mechanism contributes much less for the same points in the available parameter space, thus widening it.

The operative exploitation of the so-obtained high-temperature rates  requires some approximations to compute the dark matter energy density. To this end, one needs to follow 
DM production over the full temperature range, down to $T \siml M$. At this stage, the collinear approximation fails and the particles cannot be treated as massless anymore. As the temperature decreases due to the Universe expansion, the production process becomes increasingly sensitive to the in-vacuum mass scales $M_\eta$ and $M$. It is at this point that our treatment acquires a more phenomenological character: we follow a prescription that allows us to progressively switch off the high-temperature rates, and replace them with their Born counterpart (though still keeping the effect of thermal masses in it, see Eq.\eqref{sub_pres_2}). As for the $2 \to 2$ scattering processes we implement the switch-off with the same susceptibility factor \eqref{defkappa}. This procedure, despite being open to be improved if we knew the NLO relativistic production rate, effectively switches off
the LPM and $2\to2$ scatterings for $T \siml M_\eta$. We find that the overall (LPM and $2 \to2$) high-temperature contributions to dark-matter production give $\mathcal{O}(10)$ (20\%) corrections  for $\Delta M /M =0.1$ ($\Delta M /M =10$) to the in-vacuum Born production rate, which significantly underestimates the energy density. The effect is even more pronounced if the comparison is made with respect to the Born rate with thermal masses included.

The present work suggests to rethink freeze-in dark matter with renormalizable operators. There, the bulk of DM population is 
generally thought to be produced at low temperatures, typically around the mass of the heaviest particle participating in the
production, with the model-independent assumption that the FIMP is produced in decays or multi-particle collisions of equilibrated states in the thermal bath. However, the interactions responsible for the equilibrium of such states, either SM gauge interactions or those of some hidden sector, are the very same that induce the high-temperature processes, which are responsible for a sizeable DM production well above the ``infra-red'' domain.

Our findings trigger future investigations in many directions. 
The first, rather obvious one, is to quantitatively inspect how relevant the high-temperature production is for other dark matter models.
In the realm of the so-called simplified models, we can at least consider (i) a scalar dark matter accompanied by a colored fermion mediator interacting with quarks \cite{An:2013xka,Biondini:2019int,Arina:2020udz}; (ii) a fermionic (scalar) dark matter where the scalar (fermionic) mediator interacts with a SM lepton rather than a quark \cite{Hisano:2011cs,DiFranzo:2013vra,Garny:2015wea,Junius:2019dci}; (iii) any model where the mother particle, the heavier unstable state in the dark sector, shares interactions with the visible or hidden sector (see ref.\cite{Bernal:2017kxu} for an extensive survey of FIMP models). Another interesting aspect is the dark matter production for temperature smaller than the electroweak crossover. As briefly discussed in Sec.~\ref{sec:bench_Ps}, the accumulation of dark particles may continue for $T < 160$ GeV for $M_\eta <2$ TeV, and therefore, a careful treatment of the SM  broken phase must be addressed to handle the appearance of additional mass scales (see \cite{Ghiglieri:2016xye} for Majorana neutrinos and leptogenesis). This is most likely realized when the mediator couples to leptons, since experimental (collider) constraints are weaker and masses modestly above the electroweak scale $M_\mathrm{med} \simg 100$ GeV are allowed \cite{Garny:2015wea,Aad:2014vma,Khachatryan:2014qwa}.  Next and last, it  would be desirable, though technically very challenging, to tackle the NLO dark matter production in the relativistic regime $T \sim M$, when an additional (in-vacuum) massive state enters the dark-matter self energy. On top of being interesting/relevant on its own, this would enable a more systematic and correct matching of the two temperature regimes $T \gg M$ and $T \siml M$, and the corresponding accurate extraction of the production rate $\textrm{Im}\Pi_R$ for \textit{all} temperatures.  

\section*{Acknowledgements}
The work of S.B. is supported by the Swiss National Science Foundation (SNF) under the Ambizione grant PZ00P2\_185783. J.G. acknowledges
support by a PULSAR grant from the R\'egion Pays de la Loire.

\appendix
\numberwithin{equation}{section}

\setcounter{equation}{0}

\section{Model running couplings}
\label{sec:running}
In this work we restrict the treatment at temperatures larger than the electroweak crossover, $T > T_{c} \approx 160$ GeV. The couplings are then evaluated at the thermal scale $\mu = \pi T $ that makes it larger than any mass scale in the SM. Therefore, the three generations of quarks and leptons contribute to the running of the relevant gauge couplings, namely $g_3$ of SU(3) and $g_1$ of U(1)$_{Y}$. We parametrize the $\overline{\hbox{MS}}$  scale $\bar{\mu}$ through $t=\ln\bar{\mu}^2$ , and the RGEs for the couplings read at one loop \cite{Laine:2013lka,Biondini:2018pwp} (we neglect the contribution from the Yukawa coupling $y \ll1$ in the following equations and the evolution equation for $y$ as well) 
\begin{eqnarray}
\partial_t g^2_3 &=& \frac{g_3^4}{(4 \pi)^2} \left[ \frac{4}{3}n_G + \frac{n_S}{6} - \frac{11}{3} N_c \right] \, ,
\label{g3_beta}
\\
\partial_t g^2_2 &=& \frac{g_2^4}{(4 \pi)^2} \left[ \frac{4}{3}n_G + \frac{n_W}{6} - \frac{22}{3}  \right] \, ,
\label{g2_beta}
\\
\partial_t g^2_1 &=& \frac{g_1^4}{(4 \pi)^2} \left[ \frac{20}{9}n_G + \frac{n_W}{6} + \frac{n_S }{3} |Y_q|^2 N_c \right] \, ,
\label{g1_beta}
\\
\partial_t |h_q|^2 &=& \frac{|h_q|^2}{(4 \pi)^2} \left[  \frac{9}{2} |h_q|^2 -\frac{17}{12} g_1^2 - \frac{9}{4} g_2^2 - 6 C_F g_3^2 \right] \, ,
\label{ht_beta}
\\
\partial_t \lambda_1 &=&   \frac{1}{(4 \pi)^2}  \left\lbrace  \left[  12 \lambda_1 - \frac{3}{2} (g_1^2+3g_2^2) +2 |h_q|^2 N_c ) \right] \lambda_1 +\frac{\lambda_3^2 N_c}{2} \right.
\nonumber \\
&&\left. \hspace{1.5 cm} + \frac{3}{16} (g_1^4 + 2g_1^2 g_2^2 +3 g_2^2) -|h_q|^4 N_c  \right\rbrace \, ,
\label{lambda1_beta}
\\
\partial_t \lambda_2 &=&  \frac{1}{(4 \pi)^2}  \left\lbrace  \left[  2 \lambda_2 (N_c + 4)  - 6 g_3^2 C_F -6 Y_q^2 g_1^2\right] \lambda_2  + \lambda_3^2  \right.
\nonumber \\
&& \hspace{1.5 cm }\left. + \frac{(N_c^3 +N_c^2 -4N_c +2)}{8 N_c^2} g_3^4  + \frac{3}{4}Y_q^2 g_1^4   \right\rbrace \, ,
\label{lambda2_beta}
\\
\partial_t \lambda_3 &=& \frac{1}{(4 \pi)^2}  \left\lbrace \left[  6 \lambda_1 +2 \lambda_2 (N_c+1) + 2 \lambda_3 - \frac{3}{4} (g_1^2+3g_2^2)-3g_1^2 Y_q^2  -3 C_F g_3^2 + |h_q|^2 N_c ) \right] \lambda_3  \right.
\nonumber \\
&&  \hspace{1.5 cm } \left. +  \frac{3}{8}Y_q^2 g_1^4 +  \frac{3}{4} Y_q g_1^4 \right\rbrace \, ,
\label{lambda3_beta}
\end{eqnarray} 
where $n_G$ is the number of fermionic generations, $n_W$ the number of SU(2)$_L$ scalar doublets (the SM Higgs field) and $n_S$ the number of SU($N_c$) interacting scalar, here QCD strongly interacting scalars with $N_c=3$.
\newpage
\bibliographystyle{hieeetr}
\bibliography{draft.bib}

\begin{thebibliography}{10}

\bibitem{Bertone:2016nfn}
G.~Bertone and D.~Hooper, ``{History of dark matter},'' {\em Rev. Mod. Phys.},
  vol.~90, no.~4, p.~045002, 2018, 1605.04909.

\bibitem{Arcadi:2017kky}
G.~Arcadi, M.~Dutra, P.~Ghosh, M.~Lindner, Y.~Mambrini, M.~Pierre, S.~Profumo,
  and F.~S. Queiroz, ``{The waning of the WIMP? A review of models, searches,
  and constraints},'' {\em Eur. Phys. J. C}, vol.~78, no.~3, p.~203, 2018,
  1703.07364.

\bibitem{McDonald:2001vt}
J.~McDonald, ``{Thermally generated gauge singlet scalars as selfinteracting
  dark matter},'' {\em Phys. Rev. Lett.}, vol.~88, p.~091304, 2002,
  hep-ph/0106249.

\bibitem{Hall:2009bx}
L.~J. Hall, K.~Jedamzik, J.~March-Russell, and S.~M. West, ``{Freeze-In
  Production of FIMP Dark Matter},'' {\em JHEP}, vol.~03, p.~080, 2010,
  0911.1120.

\bibitem{Baer:2014eja}
H.~Baer, K.-Y. Choi, J.~E. Kim, and L.~Roszkowski, ``{Dark matter production in
  the early Universe: beyond the thermal WIMP paradigm},'' {\em Phys. Rept.},
  vol.~555, pp.~1--60, 2015, 1407.0017.

\bibitem{Bernal:2017kxu}
N.~Bernal, M.~Heikinheimo, T.~Tenkanen, K.~Tuominen, and V.~Vaskonen, ``{The
  Dawn of FIMP Dark Matter: A Review of Models and Constraints},'' {\em Int. J.
  Mod. Phys.}, vol.~A32, no.~27, p.~1730023, 2017, 1706.07442.

\bibitem{Yaguna:2011ei}
C.~E. Yaguna, ``{An intermediate framework between WIMP, FIMP, and EWIP dark
  matter},'' {\em JCAP}, vol.~02, p.~006, 2012, 1111.6831.

\bibitem{Krauss:2013wfa}
M.~B. Krauss, S.~Morisi, W.~Porod, and W.~Winter, ``{Higher Dimensional
  Effective Operators for Direct Dark Matter Detection},'' {\em JHEP}, vol.~02,
  p.~056, 2014, 1312.0009.

\bibitem{Elahi:2014fsa}
F.~Elahi, C.~Kolda, and J.~Unwin, ``{UltraViolet Freeze-in},'' {\em JHEP},
  vol.~03, p.~048, 2015, 1410.6157.

\bibitem{Belanger:2018ccd}
G.~B\'elanger, F.~Boudjema, A.~Goudelis, A.~Pukhov, and B.~Zaldivar,
  ``{micrOMEGAs5.0 : Freeze-in},'' {\em Comput. Phys. Commun.}, vol.~231,
  pp.~173--186, 2018, 1801.03509.

\bibitem{Lebedev:2019ton}
O.~Lebedev and T.~Toma, ``{Relativistic Freeze-in},'' {\em Phys. Lett. B},
  vol.~798, p.~134961, 2019, 1908.05491.

\bibitem{Bandyopadhyay:2020ufc}
P.~Bandyopadhyay, M.~Mitra, and A.~Roy, ``{Relativistic Freeze-in with Scalar
  Dark Matter in a Gauged $B-L$ Model and Electroweak Symmetry Breaking},'' 12
  2020, 2012.07142.

\bibitem{Baker:2016xzo}
M.~J. Baker and J.~Kopp, ``{Dark Matter Decay between Phase Transitions at the
  Weak Scale},'' {\em Phys. Rev. Lett.}, vol.~119, no.~6, p.~061801, 2017,
  1608.07578.

\bibitem{Baker:2017zwx}
M.~J. Baker, M.~Breitbach, J.~Kopp, and L.~Mittnacht, ``{Dynamic Freeze-In:
  Impact of Thermal Masses and Cosmological Phase Transitions on Dark Matter
  Production},'' {\em JHEP}, vol.~03, p.~114, 2018, 1712.03962.

\bibitem{Dvorkin:2019zdi}
C.~Dvorkin, T.~Lin, and K.~Schutz, ``{Making dark matter out of light:
  freeze-in from plasma effects},'' {\em Phys. Rev.}, vol.~D99, no.~11,
  p.~115009, 2019, 1902.08623.

\bibitem{Darme:2019wpd}
L.~Darm\'e, A.~Hryczuk, D.~Karamitros, and L.~Roszkowski, ``{Forbidden
  frozen-in dark matter},'' {\em JHEP}, vol.~11, p.~159, 2019, 1908.05685.

\bibitem{Landau:1953gr}
L.~Landau and I.~Pomeranchuk, ``{Electron cascade process at very
  high-energies},'' {\em Dokl.Akad.Nauk Ser.Fiz.}, vol.~92, pp.~735--738, 1953.

\bibitem{Landau:1953um}
L.~Landau and I.~Pomeranchuk, ``{Limits of applicability of the theory of
  bremsstrahlung electrons and pair production at high-energies},'' {\em
  Dokl.Akad.Nauk Ser.Fiz.}, vol.~92, pp.~535--536, 1953.

\bibitem{Migdal:1956tc}
A.~B. Migdal, ``{Bremsstrahlung and pair production in condensed media at
  high-energies},'' {\em Phys.Rev.}, vol.~103, pp.~1811--1820, 1956.

\bibitem{Anisimov:2010gy}
A.~Anisimov, D.~Besak, and D.~Bodeker, ``{Thermal production of relativistic
  Majorana neutrinos: Strong enhancement by multiple soft scattering},'' {\em
  JCAP}, vol.~1103, p.~042, 2011, 1012.3784.

\bibitem{Besak:2012qm}
D.~Besak and D.~Bodeker, ``{Thermal production of ultrarelativistic
  right-handed neutrinos: Complete leading-order results},'' {\em JCAP},
  vol.~1203, p.~029, 2012, 1202.1288.

\bibitem{Ghisoiu:2014mha}
I.~Ghisoiu and M.~Laine, ``{Interpolation of hard and soft dilepton rates},''
  {\em JHEP}, vol.~10, p.~083, 2014, 1407.7955.

\bibitem{Ghiglieri:2016xye}
J.~Ghiglieri and M.~Laine, ``{Neutrino dynamics below the electroweak
  crossover},'' {\em JCAP}, vol.~1607, no.~07, p.~015, 2016, 1605.07720.

\bibitem{Hisano:2011cs}
J.~Hisano, K.~Ishiwata, N.~Nagata, and T.~Takesako, ``{Direct Detection of
  Electroweak-Interacting Dark Matter},'' {\em JHEP}, vol.~07, p.~005, 2011,
  1104.0228.

\bibitem{DiFranzo:2013vra}
A.~DiFranzo, K.~I. Nagao, A.~Rajaraman, and T.~M. Tait, ``{Simplified Models
  for Dark Matter Interacting with Quarks},'' {\em JHEP}, vol.~11, p.~014,
  2013, 1308.2679.
\newblock [Erratum: JHEP 01, 162 (2014)].

\bibitem{An:2013xka}
H.~An, L.-T. Wang, and H.~Zhang, ``{Dark matter with $t$-channel mediator: a
  simple step beyond contact interaction},'' {\em Phys. Rev. D}, vol.~89,
  no.~11, p.~115014, 2014, 1308.0592.

\bibitem{Garny:2015wea}
M.~Garny, A.~Ibarra, and S.~Vogl, ``{Signatures of Majorana dark matter with
  t-channel mediators},'' {\em Int. J. Mod. Phys.}, vol.~D24, no.~07,
  p.~1530019, 2015, 1503.01500.

\bibitem{Arina:2020udz}
C.~Arina, B.~Fuks, and L.~Mantani, ``{A universal framework for t-channel dark
  matter models},'' {\em Eur. Phys. J. C}, vol.~80, no.~5, p.~409, 2020,
  2001.05024.

\bibitem{Garny:2017rxs}
M.~Garny, J.~Heisig, B.~Lülf, and S.~Vogl, ``{Coannihilation without chemical
  equilibrium},'' {\em Phys. Rev.}, vol.~D96, no.~10, p.~103521, 2017,
  1705.09292.

\bibitem{Belanger:2018sti}
G.~Bélanger {\em et~al.}, ``{LHC-friendly minimal freeze-in models},'' {\em
  JHEP}, vol.~02, p.~186, 2019, 1811.05478.

\bibitem{Garny:2018icg}
M.~Garny, J.~Heisig, M.~Hufnagel, and B.~Lülf, ``{Top-philic dark matter
  within and beyond the WIMP paradigm},'' {\em Phys. Rev.}, vol.~D97, no.~7,
  p.~075002, 2018, 1802.00814.

\bibitem{Garny:2018ali}
M.~Garny and J.~Heisig, ``{Interplay of super-WIMP and freeze-in production of
  dark matter},'' {\em Phys. Rev.}, vol.~D98, no.~9, p.~095031, 2018,
  1809.10135.

\bibitem{Junius:2019dci}
S.~Junius, L.~Lopez-Honorez, and A.~Mariotti, ``{A feeble window on leptophilic
  dark matter},'' {\em JHEP}, vol.~07, p.~136, 2019, 1904.07513.

\bibitem{Co:2015pka}
R.~T. Co, F.~D'Eramo, L.~J. Hall, and D.~Pappadopulo, ``{Freeze-In Dark Matter
  with Displaced Signatures at Colliders},'' {\em JCAP}, vol.~1512, no.~12,
  p.~024, 2015, 1506.07532.

\bibitem{Hessler:2016kwm}
A.~G. Hessler, A.~Ibarra, E.~Molinaro, and S.~Vogl, ``{Probing the scotogenic
  FIMP at the LHC},'' {\em JHEP}, vol.~01, p.~100, 2017, 1611.09540.

\bibitem{Davoli:2017swj}
A.~Davoli, A.~De~Simone, T.~Jacques, and V.~Sanz, ``{Displaced Vertices from
  Pseudo-Dirac Dark Matter},'' {\em JHEP}, vol.~11, p.~025, 2017, 1706.08985.

\bibitem{Sirunyan:2018ldc}
A.~M. Sirunyan {\em et~al.}, ``{Search for disappearing tracks as a signature
  of new long-lived particles in proton-proton collisions at $\sqrt{s} =$ 13
  TeV},'' {\em JHEP}, vol.~08, p.~016, 2018, 1804.07321.

\bibitem{Aaboud:2019trc}
M.~Aaboud {\em et~al.}, ``{Search for heavy charged long-lived particles in the
  ATLAS detector in 36.1 fb$^{-1}$ of proton-proton collision data at $\sqrt{s}
  = 13$ TeV},'' {\em Phys. Rev. D}, vol.~99, no.~9, p.~092007, 2019,
  1902.01636.

\bibitem{Alimena:2019zri}
J.~Alimena {\em et~al.}, ``{Searching for long-lived particles beyond the
  Standard Model at the Large Hadron Collider},'' {\em J. Phys. G}, vol.~47,
  no.~9, p.~090501, 2020, 1903.04497.

\bibitem{Feng:2003xh}
J.~L. Feng, A.~Rajaraman, and F.~Takayama, ``{Superweakly interacting massive
  particles},'' {\em Phys. Rev. Lett.}, vol.~91, p.~011302, 2003,
  hep-ph/0302215.

\bibitem{Feng:2003uy}
J.~L. Feng, A.~Rajaraman, and F.~Takayama, ``{SuperWIMP dark matter signals
  from the early universe},'' {\em Phys. Rev. D}, vol.~68, p.~063504, 2003,
  hep-ph/0306024.

\bibitem{Carena:1996wj}
M.~Carena, M.~Quiros, and C.~Wagner, ``{Opening the window for electroweak
  baryogenesis},'' {\em Phys. Lett. B}, vol.~380, pp.~81--91, 1996,
  hep-ph/9603420.

\bibitem{Delepine:1996vn}
D.~Delepine, J.~Gerard, R.~Gonzalez~Felipe, and J.~Weyers, ``{A Light stop and
  electroweak baryogenesis},'' {\em Phys. Lett. B}, vol.~386, pp.~183--188,
  1996, hep-ph/9604440.

\bibitem{Cline:1996cr}
J.~M. Cline and K.~Kainulainen, ``{Supersymmetric electroweak phase transition:
  Beyond perturbation theory},'' {\em Nucl. Phys. B}, vol.~482, pp.~73--91,
  1996, hep-ph/9605235.

\bibitem{Losada:1996ju}
M.~Losada, ``{High temperature dimensional reduction of the MSSM and other
  multiscalar models},'' {\em Phys. Rev. D}, vol.~56, pp.~2893--2913, 1997,
  hep-ph/9605266.

\bibitem{Laine:1996ms}
M.~Laine, ``{Effective theories of MSSM at high temperature},'' {\em Nucl.
  Phys. B}, vol.~481, pp.~43--84, 1996, hep-ph/9605283.
\newblock [Erratum: Nucl.Phys.B 548, 637--638 (1999)].

\bibitem{Asaka:2006rw}
T.~Asaka, M.~Laine, and M.~Shaposhnikov, ``{On the hadronic contribution to
  sterile neutrino production},'' {\em JHEP}, vol.~06, p.~053, 2006,
  hep-ph/0605209.

\bibitem{Bodeker:2015exa}
D.~B\"odeker, M.~Sangel, and M.~W\"ormann, ``{Equilibration, particle
  production, and self-energy},'' {\em Phys. Rev. D}, vol.~93, no.~4,
  p.~045028, 2016, 1510.06742.

\bibitem{Laine:2016hma}
M.~Laine and A.~Vuorinen, ``{Basics of Thermal Field Theory},'' {\em Lect.
  Notes Phys.}, vol.~925, pp.~pp.1--281, 2016, 1701.01554.

\bibitem{Laine:2011pq}
M.~Laine and Y.~Schroder, ``{Thermal right-handed neutrino production rate in
  the non-relativistic regime},'' {\em JHEP}, vol.~02, p.~068, 2012, 1112.1205.

\bibitem{Ghisoiu:2014ena}
I.~Ghisoiu and M.~Laine, ``{Right-handed neutrino production rate at $T > 160$
  GeV},'' {\em JCAP}, vol.~12, p.~032, 2014, 1411.1765.

\bibitem{Braaten:1989mz}
E.~Braaten and R.~D. Pisarski, ``{Soft Amplitudes in Hot Gauge Theories: A
  General Analysis},'' {\em Nucl.Phys.}, vol.~B337, p.~569, 1990.

\bibitem{Frenkel:1989br}
J.~Frenkel and J.~Taylor, ``{High Temperature Limit of Thermal QCD},'' {\em
  Nucl.Phys.}, vol.~B334, p.~199, 1990.

\bibitem{Taylor:1990ia}
J.~C. Taylor and S.~M.~H. Wong, ``{The Effective Action of Hard Thermal Loops
  in {QCD}},'' {\em Nucl. Phys.}, vol.~B346, pp.~115--128, 1990.

\bibitem{Frenkel:1991ts}
J.~Frenkel and J.~C. Taylor, ``{Hard thermal QCD, forward scattering and
  effective actions},'' {\em Nucl. Phys.}, vol.~B374, pp.~156--168, 1992.

\bibitem{Braaten:1991gm}
E.~Braaten and R.~D. Pisarski, ``{Simple effective Lagrangian for hard thermal
  loops},'' {\em Phys. Rev.}, vol.~D45, pp.~1827--1830, 1992.

\bibitem{Biondini:2018pwp}
S.~Biondini and M.~Laine, ``{Thermal dark matter co-annihilating with a
  strongly interacting scalar},'' {\em JHEP}, vol.~04, p.~072, 2018,
  1801.05821.

\bibitem{Ghiglieri:2020dpq}
J.~Ghiglieri, A.~Kurkela, M.~Strickland, and A.~Vuorinen, ``{Perturbative
  Thermal QCD: Formalism and Applications},'' {\em Phys. Rept.}, vol.~880,
  pp.~1--73, 2020, 2002.10188.

\bibitem{Ghiglieri:2014kma}
J.~Ghiglieri and G.~D. Moore, ``{Low Mass Thermal Dilepton Production at NLO in
  a Weakly Coupled Quark-Gluon Plasma},'' {\em JHEP}, vol.~12, p.~029, 2014,
  1410.4203.

\bibitem{Strassler:1990nw}
M.~J. Strassler and M.~E. Peskin, ``{The Heavy top quark threshold: QCD and the
  Higgs},'' {\em Phys. Rev. D}, vol.~43, pp.~1500--1514, 1991.

\bibitem{Laine:2013vpa}
M.~Laine, ``{Thermal 2-loop master spectral function at finite momentum},''
  {\em JHEP}, vol.~05, p.~083, 2013, 1304.0202.

\bibitem{Laine:2013lka}
M.~Laine, ``{Thermal right-handed neutrino production rate in the relativistic
  regime},'' {\em JHEP}, vol.~08, p.~138, 2013, 1307.4909.

\bibitem{Laine:2013vma}
M.~Laine, ``{NLO thermal dilepton rate at non-zero momentum},'' {\em JHEP},
  vol.~1311, p.~120, 2013, 1310.0164.

\bibitem{Jackson:2019mop}
G.~Jackson, ``{Two-loop thermal spectral functions with general kinematics},''
  {\em Phys. Rev. D}, vol.~100, no.~11, p.~116019, 2019, 1910.07552.

\bibitem{Ghiglieri:2020mhm}
J.~Ghiglieri, G.~Jackson, M.~Laine, and Y.~Zhu, ``{Gravitational wave
  background from Standard Model physics: Complete leading order},'' {\em
  JHEP}, vol.~07, p.~092, 2020, 2004.11392.

\bibitem{Arcadi:2013aba}
G.~Arcadi and L.~Covi, ``{Minimal Decaying Dark Matter and the LHC},'' {\em
  JCAP}, vol.~08, p.~005, 2013, 1305.6587.

\bibitem{Aghanim:2018eyx}
N.~Aghanim {\em et~al.}, ``{Planck 2018 results. VI. Cosmological
  parameters},'' {\em Astron. Astrophys.}, vol.~641, p.~A6, 2020, 1807.06209.

\bibitem{Edsjo:1997bg}
J.~Edsjo and P.~Gondolo, ``{Neutralino relic density including
  coannihilations},'' {\em Phys. Rev. D}, vol.~56, pp.~1879--1894, 1997,
  hep-ph/9704361.

\bibitem{deSimone:2014pda}
A.~De~Simone, G.~F. Giudice, and A.~Strumia, ``{Benchmarks for Dark Matter
  Searches at the LHC},'' {\em JHEP}, vol.~06, p.~081, 2014, 1402.6287.

\bibitem{Ellis:2014ipa}
J.~Ellis, K.~A. Olive, and J.~Zheng, ``{The Extent of the Stop Coannihilation
  Strip},'' {\em Eur. Phys. J. C}, vol.~74, p.~2947, 2014, 1404.5571.

\bibitem{Liew:2016hqo}
S.~P. Liew and F.~Luo, ``{Effects of QCD bound states on dark matter relic
  abundance},'' {\em JHEP}, vol.~02, p.~091, 2017, 1611.08133.

\bibitem{Kim:2016kxt}
S.~Kim and M.~Laine, ``{On thermal corrections to near-threshold
  annihilation},'' {\em JCAP}, vol.~01, p.~013, 2017, 1609.00474.

\bibitem{Mitridate:2017izz}
A.~Mitridate, M.~Redi, J.~Smirnov, and A.~Strumia, ``{Cosmological Implications
  of Dark Matter Bound States},'' {\em JCAP}, vol.~05, p.~006, 2017,
  1702.01141.

\bibitem{Biondini:2018ovz}
S.~Biondini and S.~Vogl, ``{Coloured coannihilations: Dark matter phenomenology
  meets non-relativistic EFTs},'' {\em JHEP}, vol.~02, p.~016, 2019,
  1811.02581.

\bibitem{Harz:2018csl}
J.~Harz and K.~Petraki, ``{Radiative bound-state formation in unbroken
  perturbative non-Abelian theories and implications for dark matter},'' {\em
  JHEP}, vol.~07, p.~096, 2018, 1805.01200.

\bibitem{Bodwin:1994jh}
G.~T. Bodwin, E.~Braaten, and G.~Lepage, ``{Rigorous QCD analysis of inclusive
  annihilation and production of heavy quarkonium},'' {\em Phys. Rev. D},
  vol.~51, pp.~1125--1171, 1995, hep-ph/9407339.
\newblock [Erratum: Phys.Rev.D 55, 5853 (1997)].

\bibitem{Kim:2016zyy}
S.~Kim and M.~Laine, ``{Rapid thermal co-annihilation through bound states in
  QCD},'' {\em JHEP}, vol.~07, p.~143, 2016, 1602.08105.

\bibitem{Matsui:1986dk}
T.~Matsui and H.~Satz, ``{$J/\psi$ Suppression by Quark-Gluon Plasma
  Formation},'' {\em Phys. Lett.}, vol.~B178, pp.~416--422, 1986.

\bibitem{Mocsy:2013syh}
A.~Mocsy, P.~Petreczky, and M.~Strickland, ``{Quarkonia in the Quark Gluon
  Plasma},'' {\em Int. J. Mod. Phys.}, vol.~A28, p.~1340012, 2013, 1302.2180.

\bibitem{Rothkopf:2019ipj}
A.~Rothkopf, ``{Heavy Quarkonium in Extreme Conditions},'' {\em Phys. Rept.},
  vol.~858, pp.~1--117, 2020, 1912.02253.

\bibitem{Laine:2006ns}
M.~Laine, O.~Philipsen, P.~Romatschke, and M.~Tassler, ``{Real-time static
  potential in hot QCD},'' {\em JHEP}, vol.~03, p.~054, 2007, hep-ph/0611300.

\bibitem{Beraudo:2007ky}
A.~Beraudo, J.~P. Blaizot, and C.~Ratti, ``{Real and imaginary-time Q anti-Q
  correlators in a thermal medium},'' {\em Nucl. Phys.}, vol.~A806,
  pp.~312--338, 2008, 0712.4394.

\bibitem{Brambilla:2008cx}
N.~Brambilla, J.~Ghiglieri, A.~Vairo, and P.~Petreczky, ``{Static
  quark-antiquark pairs at finite temperature},'' {\em Phys. Rev.}, vol.~D78,
  p.~014017, 2008, 0804.0993.

\bibitem{Biondini:2017ufr}
S.~Biondini and M.~Laine, ``{Re-derived overclosure bound for the inert doublet
  model},'' {\em JHEP}, vol.~08, p.~047, 2017, 1706.01894.

\bibitem{Binder:2018znk}
T.~Binder, L.~Covi, and K.~Mukaida, ``{Dark Matter Sommerfeld-enhanced
  annihilation and Bound-state decay at finite temperature},'' {\em Phys.
  Rev.}, vol.~D98, no.~11, p.~115023, 2018, 1808.06472.

\bibitem{Biondini:2019zdo}
S.~Biondini, S.~Kim, and M.~Laine, ``{Non-relativistic susceptibility and a
  dark matter application},'' {\em JCAP}, vol.~10, p.~078, 2019, 1908.07541.

\bibitem{Laine:2015kra}
M.~Laine and M.~Meyer, ``{Standard Model thermodynamics across the electroweak
  crossover},'' {\em JCAP}, vol.~07, p.~035, 2015, 1503.04935.

\bibitem{CMS-PAS-EXO-16-036}
``{Search for heavy stable charged particles with $12.9~\mathrm{fb}^{-1}$ of
  2016 data},'' Tech. Rep. CMS-PAS-EXO-16-036, CERN, Geneva, 2016.

\bibitem{Jedamzik:2006xz}
K.~Jedamzik, ``{Big bang nucleosynthesis constraints on hadronically and
  electromagnetically decaying relic neutral particles},'' {\em Phys. Rev. D},
  vol.~74, p.~103509, 2006, hep-ph/0604251.

\bibitem{Kawasaki:2017bqm}
M.~Kawasaki, K.~Kohri, T.~Moroi, and Y.~Takaesu, ``{Revisiting Big-Bang
  Nucleosynthesis Constraints on Long-Lived Decaying Particles},'' {\em Phys.
  Rev. D}, vol.~97, no.~2, p.~023502, 2018, 1709.01211.

\bibitem{Biondini:2019int}
S.~Biondini and S.~Vogl, ``{Scalar dark matter coannihilating with a coloured
  fermion},'' {\em JHEP}, vol.~11, p.~147, 2019, 1907.05766.

\bibitem{Aad:2014vma}
G.~Aad {\em et~al.}, ``{Search for direct production of charginos, neutralinos
  and sleptons in final states with two leptons and missing transverse momentum
  in $pp$ collisions at $\sqrt{s} =$ 8 TeV with the ATLAS detector},'' {\em
  JHEP}, vol.~05, p.~071, 2014, 1403.5294.

\bibitem{Khachatryan:2014qwa}
V.~Khachatryan {\em et~al.}, ``{Searches for electroweak production of
  charginos, neutralinos, and sleptons decaying to leptons and W, Z, and Higgs
  bosons in pp collisions at 8 TeV},'' {\em Eur. Phys. J. C}, vol.~74, no.~9,
  p.~3036, 2014, 1405.7570.

\end{thebibliography}

\end{document}